\documentclass[aps,prl,reprint,balancelastpage,nofootinbib,preprintnumbers,superscriptaddress]{revtex4-1}

\usepackage{slashed}
\usepackage{bm}
\usepackage{latexsym,amssymb,amsmath,float,url,mathrsfs}
\usepackage{latexsym}
\usepackage{graphicx}
\usepackage{epstopdf}
\usepackage{amsmath}
\usepackage{comment}
\usepackage{subfigure}
\usepackage{natbib}
\usepackage{hyperref}
\usepackage{pifont}
\usepackage{blindtext}
\usepackage{adjustbox}
\usepackage{multirow}
\usepackage{tabularx}
\usepackage[table]{xcolor}
\usepackage{orcidlink}
\usepackage{tikz}
\usetikzlibrary{tikzmark}
\usepackage{fontawesome}

\usepackage{blkarray}
\usepackage{graphicx}
\usepackage{amsfonts}
\usepackage{soul}
\usepackage{amssymb}
\usepackage{amsmath}
\usepackage{cancel}
\usepackage{tcolorbox}
\usepackage{tikz-feynman}
\usepackage{tcolorbox}

\usepackage{graphics,appendix,afterpage,makecell} 

\definecolor{oucrimsonred}{rgb}{0.6, 0.0, 0.0}
\definecolor{persianblue}{rgb}{0.11, 0.22, 0.73}
\definecolor{forestgreen}{rgb}{0.13,0.35,0.13}
\definecolor{lightgray}{rgb}{0.83, 0.83, 0.83}
 \hypersetup{colorlinks, citecolor=oucrimsonred, linkcolor=black, urlcolor=oucrimsonred}
\definecolor{cornellred}{rgb}{0.7, 0.11, 0.11}
\definecolor{navyblue}{rgb}{0.0, 0.0, 0.5}
\definecolor{amethyst}{rgb}{0.6, 0.4, 0.8}
\definecolor{yellow}{rgb}{1.0, 1.0, 0.0}
\definecolor{firebrick}{rgb}{0.7, 0.13, 0.13}
\definecolor{tangerineyellow}{rgb}{1.0, 0.8, 0.0}
\definecolor{deepfuchsia}{rgb}{0.76, 0.33, 0.76}
\definecolor{amber}{rgb}{1.0, 0.75, 0.0}
\definecolor{VioletRed4}{rgb}{0.55, 0.13, .32}
\definecolor{indiagreen}{rgb}{0.07, 0.53, 0.03}
\definecolor{VioletRed4}{rgb}{0.55, 0.13, .32}
\newcommand{\be}{\begin{equation}}
\newcommand{\ee}{\end{equation}}
\newcommand{\bea}{\begin{equation} \begin{aligned}}
\newcommand{\eea}{\end{aligned} \end{equation}}

\definecolor{oucrimsonred}{rgb}{0.6, 0.0, 0.0}
\newcommand\vertarrowbox[3][6ex]{%
  \begin{array}[t]{@{}c@{}} #2 \\
  \left\uparrow\vcenter{\hrule height #1}\right.\kern-\nulldelimiterspace\\
  \makebox[0pt]{\scriptsize#3}
  \end{array}%
}

\definecolor{verdechiaro}{rgb}{0.6,1,0.6}
\definecolor{giallochiaro}{rgb}{1,1,0.6}
\definecolor{bluscuro}{rgb}{0.15, 0.2, 0.9}
\definecolor{verdes}{rgb}{0.1, 0.5, 0.1}%
\definecolor{tangerineyellow}{rgb}{1.0, 0.8, 0.0}

\definecolor{americanrose}{rgb}{1.0, 0.01, 0.24}
\definecolor{cobalt}{rgb}{0.0, 0.28, 0.67}
\definecolor{brandeisblue}{rgb}{0.0, 0.44, 1.0}
\definecolor{mycolor}{rgb}{0.0, 0.0, 0.5}
\definecolor{oxfordblue}{rgb}{0.0, 0.13, 0.28}
\definecolor{azure}{rgb}{0.0, 0.5, 1.0}
\definecolor{turquoiseblue}{rgb}{0.0, 1.0, 0.94}
\newtcolorbox{mynewbox}[1]{colback=white!5!white,colframe=azure!75!black,fonttitle=\bfseries,title=#1}
\newtcolorbox{mybox}{colback=mycolor!5!white,colframe=azure!75!black}
\newtcolorbox{mynamedbox}[1]{colback=mycolor!5!white,colframe=azure!75!black,title=#1}
\definecolor{venetianred}{rgb}{0.78, 0.03, 0.08}
\newtcolorbox{mynamedbox1}[1]{colback=venetianred!5!white,colframe=venetianred!80!black,title=#1}
\newtcolorbox{mynamedbox2}[1]{colback=azure!5!white,colframe=azure!80!black,title=#1}

\definecolor{verdes}{rgb}{0.1, 0.5, 0.1}%
\definecolor{cornellred}{rgb}{0.7, 0.11, 0.11}

\definecolor{VioletRed4}{rgb}{0.55, 0.13, .32}

\hypersetup{
     colorlinks   = true,
     citecolor    = violet,
     urlcolor     = violet,
     linkcolor    = violet}

\definecolor{rossocorsa}{rgb}{0.83, 0.0, 0.0}

\def\lsim{\mathrel{\rlap{\lower4pt\hbox{\hskip0.5pt$\sim$}}
    \raise1pt\hbox{$<$}}}         
\def\gsim{\mathrel{\rlap{\lower4pt\hbox{\hskip0.5pt$\sim$}}
    \raise1pt\hbox{$>$}}}         

\usepackage[normalem]{ulem}

\begin{document}

\title[]{Deciphering the  Instability of the Black Hole Ringdown Quasinormal Spectrum}

\author{A. Ianniccari\orcidlink{0009-0008-9885-7737}}
\affiliation{Department of Theoretical Physics and Gravitational Wave Science Center,  \\
24 quai E. Ansermet, CH-1211 Geneva 4, Switzerland}

\author{A. J. Iovino\orcidlink{0000-0002-8531-5962}}
\affiliation{Dipartimento di Fisica, ``Sapienza'' Universit\`a di Roma, Piazzale Aldo Moro 5, 00185, Roma, Italy}
\affiliation{INFN Sezione di Roma, Piazzale Aldo Moro 5, 00185, Roma, Italy}
\affiliation{Department of Theoretical Physics and Gravitational Wave Science Center,  \\
24 quai E. Ansermet, CH-1211 Geneva 4, Switzerland}

\author{A. Kehagias\orcidlink{0000-0002-9552-9366}}
\affiliation{Physics Division, National Technical University of Athens, Athens, 15780, Greece}

\author{P. Pani\orcidlink{0000-0003-4443-1761)}}
\affiliation{Dipartimento di Fisica, ``Sapienza'' Universit\`a di Roma, Piazzale Aldo Moro 5, 00185, Roma, Italy}
\affiliation{INFN Sezione di Roma, Piazzale Aldo Moro 5, 00185, Roma, Italy}

\author{G. Perna\orcidlink{0000-0002-7364-1904}}
\affiliation{Dipartimento di Fisica e Astronomia ``Galileo Galilei'', Universit\`a degli Studi di Padova, Via Marzolo 8, I-35131, Padova, Italy}
\affiliation{INFN, Sezione di Padova, Via Marzolo 8, I-35131, Padova, Italy}
\affiliation{Department of Theoretical Physics and Gravitational Wave Science Center,  \\
24 quai E. Ansermet, CH-1211 Geneva 4, Switzerland}

\author{D. Perrone\orcidlink{0000-0003-4430-4914}}
\affiliation{Department of Theoretical Physics and Gravitational Wave Science Center,  \\
24 quai E. Ansermet, CH-1211 Geneva 4, Switzerland}

\author{A. Riotto\orcidlink{0000-0001-6948-0856}}
\affiliation{Department of Theoretical Physics and Gravitational Wave Science Center,  \\
24 quai E. Ansermet, CH-1211 Geneva 4, Switzerland}


\begin{abstract}
\noindent
The spectrum of the quasinormal modes of the gravitational waves emitted during the ringdown phase  following the merger of two black holes is of primary importance in gravitational astronomy. However, the  spectrum is extremely sensitive to small disturbances of the system, thus potentially jeopardizing the predictions of the gravitational wave observables. We offer an analytical  and intuitive explanation of such an  instability and its properties based on the transfer matrix approach of quantum mechanics.  
We also give a simple interpretation of the fact that the prompt ringdown response in the time domain and the black hole greybody factor receive parametrically small corrections, thus being robust observables.
\end{abstract}

\maketitle

\noindent\textbf{Introduction --} 
The dynamics of perturbations in a relativistic theory, owing to the dissipative nature of radiative solutions,
is generically described by a non-Hermitian problem. Within linear perturbation theory the corresponding spectrum is described by quasinormal modes~(QNMs), complex eigenfrequencies with an imaginary part encoding the energy loss.
In the context of General Relativity, QNMs play a central role for black hole~(BH) spectroscopy~\cite{Dreyer:2003bv,Detweiler:1980gk,Berti:2005ys,Gossan:2011ha}, which is based on the measurement of the remnant QNMs during the final ringdown stage of a binary merger~~\cite{Vishveshwara:1970zz,Leaver:1986gd,Nollert:1998ys,Kokkotas:1999bd,Berti:2009kk,Konoplya:2011qq}.
If the remnant is a BH, General Relativity predicts that the entire QNM spectrum is solely determined by its mass and spin. 
Checking if the QNMs extracted from gravitational-wave signals match with this prediction is one of the main tests of gravity~\cite{LIGOScientific:2021sio,Berti:2015itd,Berti:2018vdi,Isi:2019aib,Franchini:2023eda}, a unique way to test the nature of the remnant~\cite{Maggio:2020jml,Maggio:2021ans,Maggio:2023fwy,Cardoso:2019rvt}, and a probe of the astrophysical environment around compact objects~\cite{Barausse:2014tra,Cardoso:2021wlq,Cardoso:2022whc,Destounis:2022obl}.

However, at variance with conservative systems, a puzzling feature of QNMs is that they are extremely sensitive to small disturbances, either in terms of background deformations or of slightly different boundary conditions.
This was first discovered by discretizing the effective potential governing linear perturbations of a BH~\cite{Nollert:1996rf} (see also~\cite{Daghigh:2020jyk}), then studied for environmental perturbations~\cite{Leung:1997was,Leung:1999iq,Barausse:2014tra,Barausse:2014pra,Cheung:2021bol,Berti:2022xfj} and for BH mimickers with near-horizon structure~\cite{Cardoso:2016rao,Cardoso:2016oxy,Cardoso:2017cqb,Abedi:2020ujo}, and finally put on a more formal ground in terms of the pseudospectrum of the non-Hermitian operator~\cite{Jaramillo:2020tuu}.
This ``spectral instability''~\cite{Jaramillo:2020tuu,Destounis:2021lum,Gasperin:2021kfv,Boyanov:2022ark,Jaramillo:2022kuv,Kyutoku:2022gbr,Sarkar:2023rhp,Destounis:2023nmb,Arean:2023ejh,Cownden:2023dam,Destounis:2023ruj,Courty:2023rxk,Boyanov:2023qqf,Cao:2024oud,Cardoso:2024mrw} implies that the BH QNMs 
might not necessarily be optimal observables. Nonetheless, the prompt ringdown phase in time domain is much less affected by deformations~\cite{Cardoso:2016rao,Cardoso:2016oxy,Cardoso:2017cqb,Berti:2022xfj,Kyutoku:2022gbr}, suggesting that spectrally unstable QNMs conspire to provide stable observables~\cite{Rosato:2024arw}.

The scope of this work is to elucidate the QNM spectral instability providing an analytical and intuitive
explanation based on the transfer matrix approach of quantum mechanics.
While grounded on general-relativistic BH perturbation theory, our approach is general and  applies to any non-Hermitian eigenvalue problem.
Henceforth we use natural units $\hbar=c=1$.

\vskip 0.5cm
\hspace{-0.80cm} 
\textbf{ The transfer matrix formalism --} We consider  the equation of motion of the tensor modes in the Schwarzschild spacetime to which a small potential bump is added away from the location of the photon ring~\cite{Barausse:2014tra} 
\begin{equation}
\label{eqs}
\left[\partial_{r_*}^2+\omega^2-V^\epsilon_\ell(r)\right] h_{\ell}(r)=0,  
\end{equation}
where 
 \be
r_*=r+r_s\ln\left|\frac{r-r_s}{r_s}\right|
 \ee
is the standard tortoise coordinate, $\omega$ is the frequency,  $\ell$ is the spherical-harmonic multipole, and $r_s$ is the Schwarzschild radius (for a spherically symmetric background, the azimuthal number is degenerate and can be set to zero without loss of generality).

The potential reads (say, for the odd parity helicity)~\cite{Cheung:2021bol}
\begin{eqnarray}
\label{eq:potentials}
V^\epsilon_\ell(r)&=&V_0(r)+\epsilon V_1(r),\nonumber\\
V_0(r)&=&\left(1-\frac{r_s}{r}\right)\left[\frac{\ell(\ell+1)}{r^2}-\frac{3r_s}{r^3}\right],\nonumber\\
V_1&=&
\frac{\epsilon}{r_s^2}{\rm sech}^2\left(\frac{r_*-c}{r_s}\right). 
\end{eqnarray}
The P\"oschl-Teller bump $V_1$ is located at a distance $c$ much larger than the location $r_g\approx 3 r_s/2$ of the peak of the main potential (where the photon ring approximately resides). It  has an amplitude ${\cal O}(\epsilon)$ and  a width ${\cal O}(r_s)\ll c$. 

The  problem of finding the effect of the small bump onto the frequency of the QNMs can be rephrased in the language of  the transfer matrix representation. \\
\begin{figure}[t]
    \centering
    \includegraphics[width=0.5\textwidth]{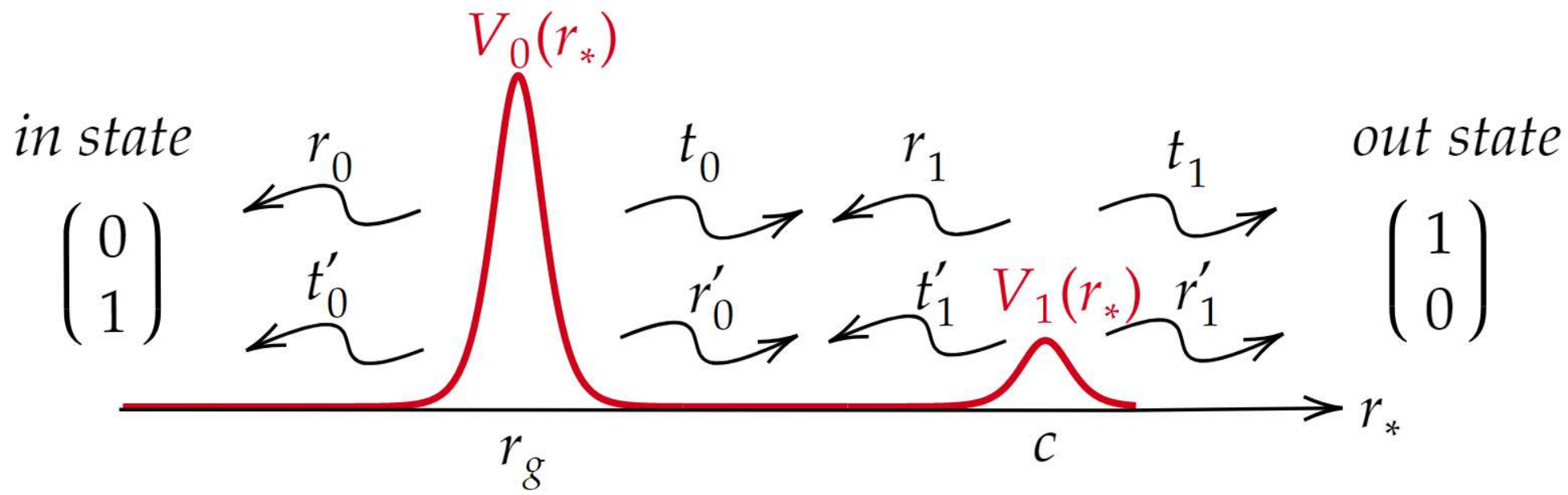}
    \caption{ 
Schematic representation of the potential with the ingoing and outgoing states and the corresponding reflection and transmission coefficients.}\label{fig:V1}
\end{figure}

We describe the ingoing and outgoing solutions  in a vector space with basis
\begin{equation}
\label{eq:basis}
   e^{-i\omega r_*} \to \begin{pmatrix} 0\\1\end{pmatrix}, \quad e^{i\omega r_*} \to \begin{pmatrix} 1\\0\end{pmatrix}.
\end{equation}
In the case of a single unspecified potential,  the  transfer matrix which maps the left basis to the right basis with respect to  a potential barrier can be generically written as (see the Supplementary Material (SM) for more details)
\begin{equation}\label{eq:transferm}
    M= \begin{pmatrix} t-r r'/t'&  r'/t'\\ -r/t'&  1/t'\end{pmatrix}.
\end{equation}
Here $r(r')$ and $t(t')$ are the reflection coefficients of a wave coming from left (right) and the transmission coefficient of a wave propagating from left to right (from right to left), respectively, see Fig.~\ref{fig:V1} and the SM.

In the language of scattering theory, the QNM frequencies are the resonances of the system and the equality
\be
\begin{pmatrix} 1\\0\end{pmatrix}=M\begin{pmatrix} 0\\1\end{pmatrix}=\begin{pmatrix} r'/t'\\1/t'\end{pmatrix}
\ee
imposes the condition $1/t'=0$, which identifies the QNMs as the poles of the transmission coefficient.

We now suppose that between the two maxima the potentials are negligible. Since both $V_0$ and $V_1$ are well localized, this assumption is certainly justified in the most relevant cases with $|c|\gg r_s$, corresponding to a model for small environmental effects away from the horizon ($c>0$)~\cite{Barausse:2014tra} or near-horizon structure ($c<0$) as those predicted by dark-matter overdensities or by certain quantum-gravity models~\cite{Cardoso:2019rvt,Bena:2022rna,Almheiri:2012rt}.

Crucially for our argument, the transfer matrix is linear. Thus, to obtain the global transfer matrix of the system of two potentials under study, it is sufficient to multiply two transfer matrices: one for potential $V_0$ and one for the  small bump $V_1$, the latter after applying the translation operator~\cite{GROSSO20001}
\be
\label{U}
U(c) = e^{i \omega  \sigma_3 c},\,\,\,  \sigma_3=\begin{pmatrix} 1& 0\\ 0&  -1\end{pmatrix}\,,
\ee
to account for the displacement of the bump relative to the main peak. Henceforth we assume the bump is on the right of the main peak (the opposite case is equivalent). The total transfer matrix reads
\begin{equation}
\label{yy}
    M_{\text{\tiny tot}} = U^{-1}(c)\cdot M_{V_1}\cdot U(c)\cdot M_{V_0},
\end{equation}
i.e. it brings a state from left to right passing through the first potential and through the second (shifted) potential.

\vskip 0.5cm
\hspace{-0.65cm} \textbf{  The effect on the QNMs --} 
Since the potentials are positive, 
 there can be no bound states  trapped in the middle
because states have always positive energy. Any wave located in the
middle will ultimately tunnel out. On the other hand, we may   expect that
the wave stays trapped for some time before it eventually
escapes. In this situation, we refer to a resonance or an unstable or metastable state. Since the QNMs are exactly such states, to study the effect on the QNMs of the tiny bumpy potential we need to identify the poles of the total transmission
coefficient. This imposes the condition
\be
\label{eq:g}
1/t'_{\text{\tiny tot}} 
=\frac{1}{t_0' t_1'}\left(1-e^{2i\omega c}r_0'r_1\right) =0\,.
\ee
The roots of Eq.~\eqref{eq:g} are no longer only  the poles of the individual transmission coefficients,  but   also those satisfying
the relation
\be
\label{eq:t_tot}
e^{-2i \omega c}=  r_0'(\omega) r_1(\omega).
\ee
This is already a signal  that the unperturbed frequencies of the QNMs may  migrate to other values. Notice that Eq.~(\ref{eq:t_tot}) is exact for any value of $\epsilon$ and $c>r_s$.

The physical interpretation of formula  (\ref{eq:g})  is more clear if we expand  $t_{\text{\tiny tot}}$ as
\begin{eqnarray}
\label{int}
t'_{\text{\tiny tot}} 
&=&t'_0\left(1+r'_0e^{2i\omega c}r_1\right.
\nonumber\\
&+&\left.r'_0e^{2i\omega c}r_1r'_0e^{2i\omega c}r_1+\cdots\right)t'_1.
\end{eqnarray}
The transmission amplitude $t_{\text{\tiny tot}}$ is given by the sum of the contributions
of all possible paths through the two potentials $V_0$ and $V_1$: on the right of the bumpy potential the wave can appear  either directly transmitted  across the two potentials from the left or after having been  reflected between the two potentials several times. 

Notice also that the 
 result is symmetric (upon changing the order of the matrices in Eq.~(\ref{yy})) for a bumpy potential located   between the horizon and the photon ring, that is for $c<0$. 

\vskip 0.5cm
\hspace{-0.4cm}\textit{Destabilization via migration of the fundamental mode -- }
To gather information from Eq.~(\ref{eq:t_tot}) we can now make the following general considerations. For sufficiently small values of $c$ we expect that the poles of the unperturbed potential are slighlty perturbed and:
\noindent
  \begin{figure*}[ht!]
    \centering
    \includegraphics[width=0.99\textwidth]{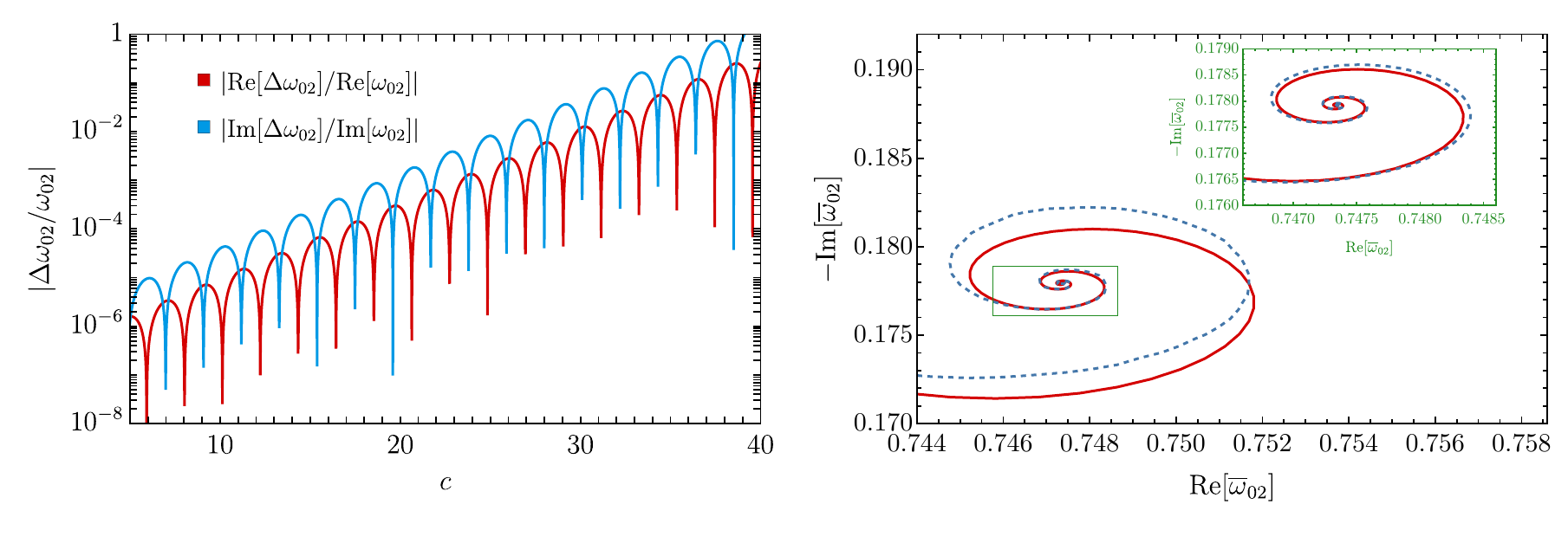}
    \caption{\textit{Left panel:} Real and imaginary parts of the migration of the frequency $\Delta\omega_{02}$ of the perturbed fundamental QNM as a function of $c$. We use units $r_s=1$.
    \textit{Right panel:} migration of the fundamental mode $\overline{\omega}_{02}$ by varying the position $c$ of the bump using the perturbation in Eq.~(\ref{mm}) in red versus the numerical results of Ref.~\cite{Cheung:2021bol} in dashed blue. The green inset shows the overlap until $c\lesssim 27$.}\label{fig:2}
\end{figure*}
\noindent
1)  $\omega c$ is  ${\cal O}(\epsilon^0)$; 
2) 
since the bump has height ${\cal O}(\epsilon)$, the reflection coefficient $r_1$ must be  ${\cal O}(\epsilon)$, that is 
\begin{equation*}
    r_1=\kappa_1\epsilon;
\end{equation*}
3) $t'_1$  is ${\cal O}(\epsilon^0 )$;
4) since $r'_0/t'_0$ is finite in this regime, the roots of Eq.~\eqref{eq:g} become the only  poles, as the poles of the unperturbed transmission coefficient are cancelled.

Since the left-handed side of Eq.~(\ref{eq:t_tot}) is finite, we infer that $r'_0$ must cancel the behaviour of $r_1$ when $\epsilon\ll1$, and therefore
\be
r'_0={\cal O}(\epsilon^{-1}).
\ee
If so, we can therefore  pick a frequency which is close to a pole  $\omega_{0}$ of $r'_0$ that generates the  fundamental QNM, i.e. Laurent expanding around the pole
\be
\label{eq:res}
r'_0\sim \frac{{\rm Res} \, r'_0 |_{\omega_{0} }}{\omega - \omega_{0}}.
\ee
Parametrising  the frequency as
\be
\overline\omega_{0} = \omega_{0} +   \Delta\omega_{0},
\ee
and substituting it in Eq.~(\ref{eq:t_tot}), we obtain at leading order in $\epsilon$
\be
\Delta\omega_{0} =  \epsilon \,e^{2i \omega_{0} c} ({\rm Res}\, r'_0)  \,\kappa_1.
\ee
Separating the real and the imaginary parts of $\omega_{0}={\rm Re} \,\omega_{0} + i\, {\rm Im} \,\omega_{0}$ (with ${\rm Im} \,\omega_{0}<0$), we obtain the shift in the frequency of the QNMs to be
\begin{eqnarray}
\label{mm}
\Delta\omega_{0}&=& \epsilon \,e^{-2c \,{\rm Im} \,\omega_{0} } \left[\cos (2c \,{\rm Re} \,\omega_{0}) + i \sin (2c\,{\rm Re} \,\omega_{0})\right]\nonumber\\
&\cdot&({\rm Res}\, r'_0)   \kappa_1.
\end{eqnarray}
In this region $\epsilon\ll |\Delta\omega_{0}/\omega_{0}|\ll 1$. This simple relation is one of the  main results of our paper. It extends the standard perturbation theory \cite{Leung:1999iq} and explains in intuitive terms why 
the correction to the 
QNM frequencies is exponentially sensitive to the distance between the two peak potentials.

To be more quantitative we can consider the potentials $V_0$ and $V_1$ and the fundamental mode 
 $(n,\ell)=(0,2)$   of the unperturbed potential with  $\omega_{02}r_s\simeq (0.74-i 0.18 )$~\cite{Kokkotas:1999bd}. The details  
for the corresponding reflection 
coefficients  are given in the SM where, for the potential $V_0$, they   are obtained in the WKB approximation.  The reflection coefficient off the P\"oschl-Teller potential  turns out to be (from now on we use units $r_s=1$)
\be
r_1= -\frac{\Gamma(1+i\omega_{02} )\Gamma(1-i\omega_{02}  - \epsilon)\Gamma(\epsilon-i\omega_{02} )}{\Gamma(1-\epsilon )\Gamma(\epsilon)\Gamma(1-i\omega_{02}  )},
\ee
which expanded for small $\epsilon$ delivers
\begin{eqnarray}
   \kappa_1&=&-\frac{i\pi} {\sinh(\pi \omega_{02} )}.
\end{eqnarray} 
  The final result is summarized in Fig.~\ref{fig:2}.  We have chosen $\epsilon=10^{-6}$ as a representative value. We recover
  the characteristic exponential migration of the fundamental mode~\cite{Barausse:2014tra,Cheung:2021bol} which however remains the $(n,\ell)=(0,2)$ mode for $c\lsim 40$. In Fig.~\ref{fig:2} we have plotted our analytical results versus the numerical ones obtained in Ref.~\cite{Cheung:2021bol}. As one can appreciate, differences are at most at the level of $\sim 10\%$ and the characteristic spiralling behavior in the complex frequency plane is well recovered.

\vskip 0.5cm
\hspace{-0.4cm}\textit{Discontinuous overtaking of the fundamental mode -- }
For large enough values of the distance between the two peaks, 
 the fundamental mode is overtaken by other longer-living modes with $n\geq 1$  ~\cite{Cheung:2021bol}.  There is a simple and intuitive explanation for such behaviour, again from Eq.~(\ref{eq:t_tot}).
 The first overtaking takes place when the imaginary part of the frequency of the $(n,\ell)=(1,2)$ mode becomes equal to  the imaginary part of the frequency of the fundamental mode $(n,\ell)=(0,2)$. Since overtones migrate faster than the fundamental mode, in this regime $\overline{\omega}_{02}\simeq\omega_{02}$, and  from  Eq.~(\ref{eq:t_tot})  the  overtaking happens roughly when  
\be
e^{-2 {\rm Im}\,\omega_{02} c}\simeq  e^{-2 {\rm Im}\,\overline\omega_{12} c}\simeq \left|r_0'(\overline\omega_{12}) r_1(\overline\omega_{12})\right|\sim 0.6\,\epsilon,
\ee 
 which gives $c\sim 38$, in good agreement with the value found in Ref.~\cite{Cheung:2021bol}. One can  proceed similarly for the higher modes and find the subsequent overtakings. 

\vskip 0.5cm
\hspace{-0.55cm}
\textit{The destabilized region --}  For even larger distances, the wave decays and loses most of its amplitude before reaching the second peak. It is efficiently reflected by the bump, slowly forming a quasi-bound state between the peaks. In such a case the frequency of the QNM cannot be anything else than of the order $1/c$, the only relevant scale in the problem. The reader can find a formal proof of such statement for the case of a two-delta model in the SM.

For the potentials of Eq.~(\ref{eq:potentials}) we can make the following estimate. Starting from Eq.~(\ref{eq:t_tot}) we know that the reflection coefficient $r_1(\omega)$ has poles with frequencies proportional to $ \epsilon$. In the regime of large values of $c$, but keeping 
\begin{equation}
    \epsilon \,c \ll 1, 
\end{equation}
 we can approximate the reflection coefficient as
\begin{equation}
    r_1 \simeq -i \frac{\epsilon}{\omega}.
\end{equation}
As for the Schwarzschild potential,  we can estimate the reflection coefficient $r_0'$  by evaluating it at $\omega \simeq  0$, 
\begin{equation}
    r_0'(\omega \simeq 0)\simeq -1.
\end{equation}
With such approximations Eq.~(\ref{eq:t_tot}) becomes
\begin{equation}
    e^{-2i\omega c} \simeq ic\,\frac{\epsilon}{\omega c}.
\end{equation}
Ignoring orders of $\mathcal{O}(\log(\omega c))$, we get the expected scaling
\begin{equation}
    \omega  \simeq \frac{i}{2c}\ln\left(i \epsilon c\right) + n\frac{\pi}{c}, \quad n \in \mathbb{N}
\end{equation}
for $c\gsim 40$ when perturbativity is lost. In particular, by considering the periodicity of the solution for the real part we obtain  
\be
{\rm Re}\,\omega_{n+1}-{\rm Re}\,\omega_n=\frac{\pi}{c},
\ee
which reproduces the result of Refs.\,\cite{Mirbabayi:2018mdm,Cheung:2021bol}.
\vskip 0.5cm
\hspace{-0.50cm} \textbf{Robust observables against the tiny perturbations --} 
After having explained the physics of the QNM change due to the addition of small perturbations, we turn our attention to discuss two observables which have been identified in the literature to be numerically robust against the presence of a small bump, again offering a simple analytical explanation of such a behaviour.

\vskip 0.5cm
\hspace{-0.50cm} \textit{Stability of the prompt ringdown in time-domain --} 
The expression (\ref{int}) explains  why the changes in the prompt ringdown signal are parametrically small and why the   fundamental QNM extracted from the observable time-domain signal is stable against small perturbations~\cite{Mirbabayi:2018mdm,Berti:2022xfj}. 

The point is that 
for times $\lsim c$, one can observe only the  wave propagating from the photon ring, which did not have time to be reflected by the tiny bump and therefore the frequency of the fundamental QNM is close to the unperturbed one. Only after a time $\gg c$, the wave seen by an asymptotic observer have time to be bounced back and forth between the two potentials, thus changing the frequency of the fundamental mode. 
To see the causality argument  more formally, one can use the retarded Green function method.  We recall that 
 time-evolution of the wave  is 
\begin{eqnarray}
 h_\ell(r_*,\tau)&=&\int_{-\infty}^{\infty}{\rm d}y \, G(r_*,\tau|y,\tau') \partial_{\tau'} h_\ell(y,\tau')\nonumber\\ 
 &-&\int_{-\infty}^{\infty}{\rm d}y \, \partial_{\tau'} G(r_*,\tau|y,\tau') h_\ell(y,\tau'),
\end{eqnarray}
where  $\tau$ indicates time and $G(r_*,\tau|y,\tau')$ is the retarded Green's function
\begin{eqnarray}
  G(r_*,\tau|y,\tau')&=&\int_{a-i\infty}^{a+i\infty}\frac{{\rm d}s}{2\pi i}\frac{e^{s(\tau-\tau')}}{W_{0+1}}\nonumber\\
  &\cdot&\left[h_\ell^-(s,r_*)h_\ell^+(s,y)\theta(r_*-y)\right.\nonumber\\
  &+&\left.h_\ell^+(s,r_*)h_\ell^-(s,y)\theta(y-r_*)
    \right],
\end{eqnarray}
where  the contour runs parallel to the $({\rm Im}\,s)$ axis, but with a positive real part ($a>0$) so that it lies to the right of all singularities, $h_\ell^{\pm}$
are the two solutions of the problem with the appropriate boundary conditions and   the   frequencies (in Laplace space) are poles  of the inverse of the  Wronskian $W_{0+1}$. For sufficiently short  times $(\tau-\tau')\ll c$, that is for large values of $({\rm Re}\,s)$, and using Ref.~\cite{Hui:2019aox}, we derive that 
\be
\frac{1}{W_{0+1}}\simeq\frac{t_0' t_1'}{2s}\sum_{k\geq 0}(-1)^k e^{-2k s c},
\ee
so that  only the first term $k=0$ survives.  Causality forbids back and forth bounces, and one observes the QNMs associated with the poles of the individual transmission coefficients. The fundamental QNM will then be very similar to that of the unperturbed system.

\vskip 0.5cm
\hspace{-0.55cm} \textit{The greybody factor --}  The greybody factor is  the absorption cross section of the BH, as it measures how much of the incident field is effectively transmitted through the potential barrier and ``absorbed" by the BH.
For real frequencies we can define the greybody factor as the absolute value squared of the transmission coefficient~\cite{Harmark:2007jy}
\be
\gamma_{\text{\tiny tot}}(\omega)=t'_{\text{\tiny tot}}(\omega)t'^*_{\text{\tiny tot}}(\omega)\,,
\ee
reducing to $\gamma_0(\omega)=t'_0(\omega)t'^*_0(\omega)$ in the unperturbed case. Using the transmission coefficients of the full potential, we get
\begin{eqnarray}
\gamma_{\text{\tiny tot}}(\omega)&=&
\frac{t'_0(\omega) t'_1(\omega)t'^*_0(\omega)t'^*_1(\omega)}{\left(1-e^{2i\omega c}r'_0r_1\right)\left(1-e^{-2i\omega c}r'^*_0r^{*}_1\right)}\,.
\end{eqnarray}

In the limit in which the destabilization via migration of the fundamental mode takes place, the pole of the transmission coefficient of the unperturbed potential cancels at leading order in $\epsilon$ (see Eq.~(\ref{eq:res})), and 
\begin{eqnarray}
\gamma_{\text{\tiny tot}}(\omega)&=&\frac{{\rm Res}\, t'_0(\omega_{02}) t'_1(\omega)t'^*_0(\omega)t'^*_1(\omega)}{(\omega-\overline\omega_{02})\left(1-e^{-2i\omega c}r'^*_0r^*_1\right)},\\
\gamma_0(\omega)&=&\frac{{\rm Res}\, t'_0(\omega_{02})}{\omega-\omega_{02}} t'^*_0(\omega).
\end{eqnarray}
The integral of the change induced by the bump, $\Delta\gamma(\omega)=\gamma_{\text{\tiny tot}}(\omega)-\gamma_0(\omega)$, over all possible frequencies 
can be computed by deforming it to a contour integral in the
complex plane,
\begin{eqnarray}
\int_0^\infty{\rm d}
\omega\Delta\gamma&=&
\frac{{\rm Res}\, t'_0(\omega_{02}) t'_1(\overline\omega_{02})t'^*_0(\overline\omega_{02})t'^*_1(\overline\omega_{02})}{\left(1-e^{-2i\overline\omega_{02}c} r'^*_0(\overline\omega_{02})r^*_1(\overline\omega_{02})\right)}\nonumber\\
&-&{\rm Res}\, t'_0(\omega_{02}) t'^*_0(\omega_{02})\,.
\end{eqnarray}
In the $\epsilon\ll1$ limit,  $t'^*_0(\overline\omega_{02})\simeq t'^*_0(\omega_{02})$,    $t'^*_1(\overline\omega_{02})\sim 1$, and $r^*_1(\overline\omega_{02})\sim \epsilon$. Therefore,  the previous expression reduces to
\begin{equation}
\int_0^\infty{\rm d}
\omega\Delta\gamma={\rm Res}\, t'_0(\omega_{02})t'^*_0(\omega_{02})\left[ t'_1(\overline\omega_{02})-1\right].
\end{equation}
 The key point is now that (see the SM)
\be
 t'_1(\overline\omega_{02})\simeq 1+{\cal O}(\epsilon)e^{-\pi{\rm Re}\,\overline\omega_{02}}.
\ee
Therefore, the integral over the frequencies, which is a finite quantity, must be  ${\cal O}(\epsilon)$ and does not feel the exponential migration of the frequency of the fundamental QNM. This confirms the findings of Refs.~\cite{Rosato:2024arw,Oshita:2024fzf}, where the greybody factor has been proposed as an observable which does not feel the QNM instability.

\vskip 0.5cm
\hspace{-0.60cm} \textbf{Conclusions --} It is well-known that  small perturbations  to the effective potential governing the gravitational-wave generation and propagation around a BH  can lead to  large variations in the QNM spectrum. In this paper we have clarified the physics of such perturbations using the simple language of the transfer matrix of quantum mechanics, reproducing the known results in the literature and providing simple and intuitive explanations for the various regimes. We have also explained the  physical behavior of the time-domain signals for tiny  perturbations in the potential and why the fundamental mode in the prompt ringdown signal is not altered by the presence of small bumps in the potential. Similarly, we have offered a simple motivation of why the greybody factor receives parametrically small corrections, thus being a robust observable against perturbations in the effective potential. 

There are other directions over which expand our work. For instance, it would be interesting to understand the case of spinning BHs as well and to identify other observables insensitive to the migration of the QNM spectrum. Furthermore, it is well-known that the QNMs
 can be thought of as  null particles trapped at the unstable circular orbit and slowly leaking out
\cite{Goebel,Ferrari:1984zz,Mashhoon:1985cya,Berti:2005eb}, with  
the real part of the QNM
frequency being set  by the angular velocity at the
unstable null geodesic, while  the imaginary part being  related to
the instability timescale of the orbit ~\cite{Cornish:2003ig,Cardoso:2008bp} and to the properties of the BH collapse \cite{Ianniccari:2024ltb}.  An interesting question is if such a correspondence persists in the presence of small perturbations of the effective potential. For instance (the reader is referred again to the SM), the quasi-bound state frequency $\omega_{02}\sim 1/c$ in the destabilized region can be reproduced by such arguments. We intend to investigate these issues in the near future.

\vskip 0.5cm
\noindent
\textit{Note Added --}
After the completion of this work,  we have become aware of a similar work by Y. Yang et al.~\cite{Yang} and have coordinated the submission to the archive.  Our results, when overlap is possible, agree with theirs.

\begin{acknowledgments}
\vspace{5pt}\noindent\emph{Acknowledgments --}
We thank E.~Berti and V.~Cardoso for comments on the draft, and V.~De Luca and A.~Tagliente for useful discussions. A.R.  acknowledges support from the  Swiss National Science Foundation (project number CRSII5\_213497).
D.~P. and A.R. are supported by the Boninchi Foundation for the project ``PBHs in the Era of GW Astronomy''.
P.P. is partially supported by the MUR PRIN Grant 2020KR4KN2 ``String Theory as a bridge between Gauge Theories and Quantum Gravity'', by the FARE programme (GW-NEXT, CUP:~B84I20000100001), and by the INFN TEONGRAV initiative.
\end{acknowledgments}

\bibliography{Draft}

\begin{thebibliography}{68}%
\makeatletter
\providecommand \@ifxundefined [1]{%
 \@ifx{#1\undefined}
}%
\providecommand \@ifnum [1]{%
 \ifnum #1\expandafter \@firstoftwo
 \else \expandafter \@secondoftwo
 \fi
}%
\providecommand \@ifx [1]{%
 \ifx #1\expandafter \@firstoftwo
 \else \expandafter \@secondoftwo
 \fi
}%
\providecommand \natexlab [1]{#1}%
\providecommand \enquote  [1]{``#1''}%
\providecommand \bibnamefont  [1]{#1}%
\providecommand \bibfnamefont [1]{#1}%
\providecommand \citenamefont [1]{#1}%
\providecommand \href@noop [0]{\@secondoftwo}%
\providecommand \href [0]{\begingroup \@sanitize@url \@href}%
\providecommand \@href[1]{\@@startlink{#1}\@@href}%
\providecommand \@@href[1]{\endgroup#1\@@endlink}%
\providecommand \@sanitize@url [0]{\catcode `\\12\catcode `\$12\catcode `\&12\catcode `\#12\catcode `\^12\catcode `\_12\catcode `\%12\relax}%
\providecommand \@@startlink[1]{}%
\providecommand \@@endlink[0]{}%
\providecommand \url  [0]{\begingroup\@sanitize@url \@url }%
\providecommand \@url [1]{\endgroup\@href {#1}{\urlprefix }}%
\providecommand \urlprefix  [0]{URL }%
\providecommand \Eprint [0]{\href }%
\providecommand \doibase [0]{http://dx.doi.org/}%
\providecommand \selectlanguage [0]{\@gobble}%
\providecommand \bibinfo  [0]{\@secondoftwo}%
\providecommand \bibfield  [0]{\@secondoftwo}%
\providecommand \translation [1]{[#1]}%
\providecommand \BibitemOpen [0]{}%
\providecommand \bibitemStop [0]{}%
\providecommand \bibitemNoStop [0]{.\EOS\space}%
\providecommand \EOS [0]{\spacefactor3000\relax}%
\providecommand \BibitemShut  [1]{\csname bibitem#1\endcsname}%
\let\auto@bib@innerbib\@empty
\bibitem [{\citenamefont {Dreyer}\ \emph {et~al.}(2004)\citenamefont {Dreyer}, \citenamefont {Kelly}, \citenamefont {Krishnan}, \citenamefont {Finn}, \citenamefont {Garrison},\ and\ \citenamefont {Lopez-Aleman}}]{Dreyer:2003bv}%
  \BibitemOpen
  \bibfield  {author} {\bibinfo {author} {\bibfnamefont {O.}~\bibnamefont {Dreyer}}, \bibinfo {author} {\bibfnamefont {B.~J.}\ \bibnamefont {Kelly}}, \bibinfo {author} {\bibfnamefont {B.}~\bibnamefont {Krishnan}}, \bibinfo {author} {\bibfnamefont {L.~S.}\ \bibnamefont {Finn}}, \bibinfo {author} {\bibfnamefont {D.}~\bibnamefont {Garrison}}, \ and\ \bibinfo {author} {\bibfnamefont {R.}~\bibnamefont {Lopez-Aleman}},\ }\href {\doibase 10.1088/0264-9381/21/4/003} {\bibfield  {journal} {\bibinfo  {journal} {Class. Quant. Grav.}\ }\textbf {\bibinfo {volume} {21}},\ \bibinfo {pages} {787} (\bibinfo {year} {2004})},\ \Eprint {http://arxiv.org/abs/gr-qc/0309007} {arXiv:gr-qc/0309007} \BibitemShut {NoStop}%
\bibitem [{\citenamefont {Detweiler}(1980)}]{Detweiler:1980gk}%
  \BibitemOpen
  \bibfield  {author} {\bibinfo {author} {\bibfnamefont {S.~L.}\ \bibnamefont {Detweiler}},\ }\href {\doibase 10.1086/158109} {\bibfield  {journal} {\bibinfo  {journal} {Astrophys. J.}\ }\textbf {\bibinfo {volume} {239}},\ \bibinfo {pages} {292} (\bibinfo {year} {1980})}\BibitemShut {NoStop}%
\bibitem [{\citenamefont {Berti}\ \emph {et~al.}(2006)\citenamefont {Berti}, \citenamefont {Cardoso},\ and\ \citenamefont {Will}}]{Berti:2005ys}%
  \BibitemOpen
  \bibfield  {author} {\bibinfo {author} {\bibfnamefont {E.}~\bibnamefont {Berti}}, \bibinfo {author} {\bibfnamefont {V.}~\bibnamefont {Cardoso}}, \ and\ \bibinfo {author} {\bibfnamefont {C.~M.}\ \bibnamefont {Will}},\ }\href {\doibase 10.1103/PhysRevD.73.064030} {\bibfield  {journal} {\bibinfo  {journal} {Phys. Rev. D}\ }\textbf {\bibinfo {volume} {73}},\ \bibinfo {pages} {064030} (\bibinfo {year} {2006})},\ \Eprint {http://arxiv.org/abs/gr-qc/0512160} {arXiv:gr-qc/0512160} \BibitemShut {NoStop}%
\bibitem [{\citenamefont {Gossan}\ \emph {et~al.}(2012)\citenamefont {Gossan}, \citenamefont {Veitch},\ and\ \citenamefont {Sathyaprakash}}]{Gossan:2011ha}%
  \BibitemOpen
  \bibfield  {author} {\bibinfo {author} {\bibfnamefont {S.}~\bibnamefont {Gossan}}, \bibinfo {author} {\bibfnamefont {J.}~\bibnamefont {Veitch}}, \ and\ \bibinfo {author} {\bibfnamefont {B.~S.}\ \bibnamefont {Sathyaprakash}},\ }\href {\doibase 10.1103/PhysRevD.85.124056} {\bibfield  {journal} {\bibinfo  {journal} {Phys. Rev. D}\ }\textbf {\bibinfo {volume} {85}},\ \bibinfo {pages} {124056} (\bibinfo {year} {2012})},\ \Eprint {http://arxiv.org/abs/1111.5819} {arXiv:1111.5819 [gr-qc]} \BibitemShut {NoStop}%
\bibitem [{\citenamefont {Vishveshwara}(1970)}]{Vishveshwara:1970zz}%
  \BibitemOpen
  \bibfield  {author} {\bibinfo {author} {\bibfnamefont {C.~V.}\ \bibnamefont {Vishveshwara}},\ }\href {\doibase 10.1038/227936a0} {\bibfield  {journal} {\bibinfo  {journal} {Nature}\ }\textbf {\bibinfo {volume} {227}},\ \bibinfo {pages} {936} (\bibinfo {year} {1970})}\BibitemShut {NoStop}%
\bibitem [{\citenamefont {Leaver}(1986)}]{Leaver:1986gd}%
  \BibitemOpen
  \bibfield  {author} {\bibinfo {author} {\bibfnamefont {E.~W.}\ \bibnamefont {Leaver}},\ }\href {\doibase 10.1103/PhysRevD.34.384} {\bibfield  {journal} {\bibinfo  {journal} {Phys. Rev. D}\ }\textbf {\bibinfo {volume} {34}},\ \bibinfo {pages} {384} (\bibinfo {year} {1986})}\BibitemShut {NoStop}%
\bibitem [{\citenamefont {Nollert}\ and\ \citenamefont {Price}(1999)}]{Nollert:1998ys}%
  \BibitemOpen
  \bibfield  {author} {\bibinfo {author} {\bibfnamefont {H.-P.}\ \bibnamefont {Nollert}}\ and\ \bibinfo {author} {\bibfnamefont {R.~H.}\ \bibnamefont {Price}},\ }\href {\doibase 10.1063/1.532698} {\bibfield  {journal} {\bibinfo  {journal} {J. Math. Phys.}\ }\textbf {\bibinfo {volume} {40}},\ \bibinfo {pages} {980} (\bibinfo {year} {1999})},\ \Eprint {http://arxiv.org/abs/gr-qc/9810074} {arXiv:gr-qc/9810074} \BibitemShut {NoStop}%
\bibitem [{\citenamefont {Kokkotas}\ and\ \citenamefont {Schmidt}(1999)}]{Kokkotas:1999bd}%
  \BibitemOpen
  \bibfield  {author} {\bibinfo {author} {\bibfnamefont {K.~D.}\ \bibnamefont {Kokkotas}}\ and\ \bibinfo {author} {\bibfnamefont {B.~G.}\ \bibnamefont {Schmidt}},\ }\href {\doibase 10.12942/lrr-1999-2} {\bibfield  {journal} {\bibinfo  {journal} {Living Rev. Rel.}\ }\textbf {\bibinfo {volume} {2}},\ \bibinfo {pages} {2} (\bibinfo {year} {1999})},\ \Eprint {http://arxiv.org/abs/gr-qc/9909058} {arXiv:gr-qc/9909058} \BibitemShut {NoStop}%
\bibitem [{\citenamefont {Berti}\ \emph {et~al.}(2009)\citenamefont {Berti}, \citenamefont {Cardoso},\ and\ \citenamefont {Starinets}}]{Berti:2009kk}%
  \BibitemOpen
  \bibfield  {author} {\bibinfo {author} {\bibfnamefont {E.}~\bibnamefont {Berti}}, \bibinfo {author} {\bibfnamefont {V.}~\bibnamefont {Cardoso}}, \ and\ \bibinfo {author} {\bibfnamefont {A.~O.}\ \bibnamefont {Starinets}},\ }\href {\doibase 10.1088/0264-9381/26/16/163001} {\bibfield  {journal} {\bibinfo  {journal} {Class. Quant. Grav.}\ }\textbf {\bibinfo {volume} {26}},\ \bibinfo {pages} {163001} (\bibinfo {year} {2009})},\ \Eprint {http://arxiv.org/abs/0905.2975} {arXiv:0905.2975 [gr-qc]} \BibitemShut {NoStop}%
\bibitem [{\citenamefont {Konoplya}\ and\ \citenamefont {Zhidenko}(2011)}]{Konoplya:2011qq}%
  \BibitemOpen
  \bibfield  {author} {\bibinfo {author} {\bibfnamefont {R.~A.}\ \bibnamefont {Konoplya}}\ and\ \bibinfo {author} {\bibfnamefont {A.}~\bibnamefont {Zhidenko}},\ }\href {\doibase 10.1103/RevModPhys.83.793} {\bibfield  {journal} {\bibinfo  {journal} {Rev. Mod. Phys.}\ }\textbf {\bibinfo {volume} {83}},\ \bibinfo {pages} {793} (\bibinfo {year} {2011})},\ \Eprint {http://arxiv.org/abs/1102.4014} {arXiv:1102.4014 [gr-qc]} \BibitemShut {NoStop}%
\bibitem [{\citenamefont {Abbott}\ \emph {et~al.}(2021)\citenamefont {Abbott} \emph {et~al.}}]{LIGOScientific:2021sio}%
  \BibitemOpen
  \bibfield  {author} {\bibinfo {author} {\bibfnamefont {R.}~\bibnamefont {Abbott}} \emph {et~al.} (\bibinfo {collaboration} {LIGO Scientific, VIRGO, KAGRA}),\ }\href@noop {} {\  (\bibinfo {year} {2021})},\ \Eprint {http://arxiv.org/abs/2112.06861} {arXiv:2112.06861 [gr-qc]} \BibitemShut {NoStop}%
\bibitem [{\citenamefont {Berti}\ \emph {et~al.}(2015)\citenamefont {Berti} \emph {et~al.}}]{Berti:2015itd}%
  \BibitemOpen
  \bibfield  {author} {\bibinfo {author} {\bibfnamefont {E.}~\bibnamefont {Berti}} \emph {et~al.},\ }\href {\doibase 10.1088/0264-9381/32/24/243001} {\bibfield  {journal} {\bibinfo  {journal} {Class. Quant. Grav.}\ }\textbf {\bibinfo {volume} {32}},\ \bibinfo {pages} {243001} (\bibinfo {year} {2015})},\ \Eprint {http://arxiv.org/abs/1501.07274} {arXiv:1501.07274 [gr-qc]} \BibitemShut {NoStop}%
\bibitem [{\citenamefont {Berti}\ \emph {et~al.}(2018)\citenamefont {Berti}, \citenamefont {Yagi}, \citenamefont {Yang},\ and\ \citenamefont {Yunes}}]{Berti:2018vdi}%
  \BibitemOpen
  \bibfield  {author} {\bibinfo {author} {\bibfnamefont {E.}~\bibnamefont {Berti}}, \bibinfo {author} {\bibfnamefont {K.}~\bibnamefont {Yagi}}, \bibinfo {author} {\bibfnamefont {H.}~\bibnamefont {Yang}}, \ and\ \bibinfo {author} {\bibfnamefont {N.}~\bibnamefont {Yunes}},\ }\href {\doibase 10.1007/s10714-018-2372-6} {\bibfield  {journal} {\bibinfo  {journal} {Gen. Rel. Grav.}\ }\textbf {\bibinfo {volume} {50}},\ \bibinfo {pages} {49} (\bibinfo {year} {2018})},\ \Eprint {http://arxiv.org/abs/1801.03587} {arXiv:1801.03587 [gr-qc]} \BibitemShut {NoStop}%
\bibitem [{\citenamefont {Isi}\ \emph {et~al.}(2019)\citenamefont {Isi}, \citenamefont {Giesler}, \citenamefont {Farr}, \citenamefont {Scheel},\ and\ \citenamefont {Teukolsky}}]{Isi:2019aib}%
  \BibitemOpen
  \bibfield  {author} {\bibinfo {author} {\bibfnamefont {M.}~\bibnamefont {Isi}}, \bibinfo {author} {\bibfnamefont {M.}~\bibnamefont {Giesler}}, \bibinfo {author} {\bibfnamefont {W.~M.}\ \bibnamefont {Farr}}, \bibinfo {author} {\bibfnamefont {M.~A.}\ \bibnamefont {Scheel}}, \ and\ \bibinfo {author} {\bibfnamefont {S.~A.}\ \bibnamefont {Teukolsky}},\ }\href {\doibase 10.1103/PhysRevLett.123.111102} {\bibfield  {journal} {\bibinfo  {journal} {Phys. Rev. Lett.}\ }\textbf {\bibinfo {volume} {123}},\ \bibinfo {pages} {111102} (\bibinfo {year} {2019})},\ \Eprint {http://arxiv.org/abs/1905.00869} {arXiv:1905.00869 [gr-qc]} \BibitemShut {NoStop}%
\bibitem [{\citenamefont {Franchini}\ and\ \citenamefont {V\"olkel}(2023)}]{Franchini:2023eda}%
  \BibitemOpen
  \bibfield  {author} {\bibinfo {author} {\bibfnamefont {N.}~\bibnamefont {Franchini}}\ and\ \bibinfo {author} {\bibfnamefont {S.~H.}\ \bibnamefont {V\"olkel}},\ }\href@noop {} {\  (\bibinfo {year} {2023})},\ \Eprint {http://arxiv.org/abs/2305.01696} {arXiv:2305.01696 [gr-qc]} \BibitemShut {NoStop}%
\bibitem [{\citenamefont {Maggio}\ \emph {et~al.}(2020)\citenamefont {Maggio}, \citenamefont {Buoninfante}, \citenamefont {Mazumdar},\ and\ \citenamefont {Pani}}]{Maggio:2020jml}%
  \BibitemOpen
  \bibfield  {author} {\bibinfo {author} {\bibfnamefont {E.}~\bibnamefont {Maggio}}, \bibinfo {author} {\bibfnamefont {L.}~\bibnamefont {Buoninfante}}, \bibinfo {author} {\bibfnamefont {A.}~\bibnamefont {Mazumdar}}, \ and\ \bibinfo {author} {\bibfnamefont {P.}~\bibnamefont {Pani}},\ }\href {\doibase 10.1103/PhysRevD.102.064053} {\bibfield  {journal} {\bibinfo  {journal} {Phys. Rev. D}\ }\textbf {\bibinfo {volume} {102}},\ \bibinfo {pages} {064053} (\bibinfo {year} {2020})},\ \Eprint {http://arxiv.org/abs/2006.14628} {arXiv:2006.14628 [gr-qc]} \BibitemShut {NoStop}%
\bibitem [{\citenamefont {Maggio}\ \emph {et~al.}(2021)\citenamefont {Maggio}, \citenamefont {Pani},\ and\ \citenamefont {Raposo}}]{Maggio:2021ans}%
  \BibitemOpen
  \bibfield  {author} {\bibinfo {author} {\bibfnamefont {E.}~\bibnamefont {Maggio}}, \bibinfo {author} {\bibfnamefont {P.}~\bibnamefont {Pani}}, \ and\ \bibinfo {author} {\bibfnamefont {G.}~\bibnamefont {Raposo}},\ }\href@noop {} {\  (\bibinfo {year} {2021})},\ \Eprint {http://arxiv.org/abs/2105.06410} {arXiv:2105.06410 [gr-qc]} \BibitemShut {NoStop}%
\bibitem [{\citenamefont {Maggio}(2023)}]{Maggio:2023fwy}%
  \BibitemOpen
  \bibfield  {author} {\bibinfo {author} {\bibfnamefont {E.}~\bibnamefont {Maggio}},\ }\href {\doibase 10.1007/978-3-031-31520-6_9} {\bibfield  {journal} {\bibinfo  {journal} {Lect. Notes Phys.}\ }\textbf {\bibinfo {volume} {1017}},\ \bibinfo {pages} {333} (\bibinfo {year} {2023})},\ \Eprint {http://arxiv.org/abs/2310.07368} {arXiv:2310.07368 [gr-qc]} \BibitemShut {NoStop}%
\bibitem [{\citenamefont {Cardoso}\ and\ \citenamefont {Pani}(2019)}]{Cardoso:2019rvt}%
  \BibitemOpen
  \bibfield  {author} {\bibinfo {author} {\bibfnamefont {V.}~\bibnamefont {Cardoso}}\ and\ \bibinfo {author} {\bibfnamefont {P.}~\bibnamefont {Pani}},\ }\href {\doibase 10.1007/s41114-019-0020-4} {\bibfield  {journal} {\bibinfo  {journal} {Living Rev. Rel.}\ }\textbf {\bibinfo {volume} {22}},\ \bibinfo {pages} {4} (\bibinfo {year} {2019})},\ \Eprint {http://arxiv.org/abs/1904.05363} {arXiv:1904.05363 [gr-qc]} \BibitemShut {NoStop}%
\bibitem [{\citenamefont {Barausse}\ \emph {et~al.}(2014)\citenamefont {Barausse}, \citenamefont {Cardoso},\ and\ \citenamefont {Pani}}]{Barausse:2014tra}%
  \BibitemOpen
  \bibfield  {author} {\bibinfo {author} {\bibfnamefont {E.}~\bibnamefont {Barausse}}, \bibinfo {author} {\bibfnamefont {V.}~\bibnamefont {Cardoso}}, \ and\ \bibinfo {author} {\bibfnamefont {P.}~\bibnamefont {Pani}},\ }\href {\doibase 10.1103/PhysRevD.89.104059} {\bibfield  {journal} {\bibinfo  {journal} {Phys. Rev. D}\ }\textbf {\bibinfo {volume} {89}},\ \bibinfo {pages} {104059} (\bibinfo {year} {2014})},\ \Eprint {http://arxiv.org/abs/1404.7149} {arXiv:1404.7149 [gr-qc]} \BibitemShut {NoStop}%
\bibitem [{\citenamefont {Cardoso}\ \emph {et~al.}(2022{\natexlab{a}})\citenamefont {Cardoso}, \citenamefont {Destounis}, \citenamefont {Duque}, \citenamefont {Macedo},\ and\ \citenamefont {Maselli}}]{Cardoso:2021wlq}%
  \BibitemOpen
  \bibfield  {author} {\bibinfo {author} {\bibfnamefont {V.}~\bibnamefont {Cardoso}}, \bibinfo {author} {\bibfnamefont {K.}~\bibnamefont {Destounis}}, \bibinfo {author} {\bibfnamefont {F.}~\bibnamefont {Duque}}, \bibinfo {author} {\bibfnamefont {R.~P.}\ \bibnamefont {Macedo}}, \ and\ \bibinfo {author} {\bibfnamefont {A.}~\bibnamefont {Maselli}},\ }\href {\doibase 10.1103/PhysRevD.105.L061501} {\bibfield  {journal} {\bibinfo  {journal} {Phys. Rev. D}\ }\textbf {\bibinfo {volume} {105}},\ \bibinfo {pages} {L061501} (\bibinfo {year} {2022}{\natexlab{a}})},\ \Eprint {http://arxiv.org/abs/2109.00005} {arXiv:2109.00005 [gr-qc]} \BibitemShut {NoStop}%
\bibitem [{\citenamefont {Cardoso}\ \emph {et~al.}(2022{\natexlab{b}})\citenamefont {Cardoso}, \citenamefont {Destounis}, \citenamefont {Duque}, \citenamefont {Panosso~Macedo},\ and\ \citenamefont {Maselli}}]{Cardoso:2022whc}%
  \BibitemOpen
  \bibfield  {author} {\bibinfo {author} {\bibfnamefont {V.}~\bibnamefont {Cardoso}}, \bibinfo {author} {\bibfnamefont {K.}~\bibnamefont {Destounis}}, \bibinfo {author} {\bibfnamefont {F.}~\bibnamefont {Duque}}, \bibinfo {author} {\bibfnamefont {R.}~\bibnamefont {Panosso~Macedo}}, \ and\ \bibinfo {author} {\bibfnamefont {A.}~\bibnamefont {Maselli}},\ }\href {\doibase 10.1103/PhysRevLett.129.241103} {\bibfield  {journal} {\bibinfo  {journal} {Phys. Rev. Lett.}\ }\textbf {\bibinfo {volume} {129}},\ \bibinfo {pages} {241103} (\bibinfo {year} {2022}{\natexlab{b}})},\ \Eprint {http://arxiv.org/abs/2210.01133} {arXiv:2210.01133 [gr-qc]} \BibitemShut {NoStop}%
\bibitem [{\citenamefont {Destounis}\ \emph {et~al.}(2023)\citenamefont {Destounis}, \citenamefont {Kulathingal}, \citenamefont {Kokkotas},\ and\ \citenamefont {Papadopoulos}}]{Destounis:2022obl}%
  \BibitemOpen
  \bibfield  {author} {\bibinfo {author} {\bibfnamefont {K.}~\bibnamefont {Destounis}}, \bibinfo {author} {\bibfnamefont {A.}~\bibnamefont {Kulathingal}}, \bibinfo {author} {\bibfnamefont {K.~D.}\ \bibnamefont {Kokkotas}}, \ and\ \bibinfo {author} {\bibfnamefont {G.~O.}\ \bibnamefont {Papadopoulos}},\ }\href {\doibase 10.1103/PhysRevD.107.084027} {\bibfield  {journal} {\bibinfo  {journal} {Phys. Rev. D}\ }\textbf {\bibinfo {volume} {107}},\ \bibinfo {pages} {084027} (\bibinfo {year} {2023})},\ \Eprint {http://arxiv.org/abs/2210.09357} {arXiv:2210.09357 [gr-qc]} \BibitemShut {NoStop}%
\bibitem [{\citenamefont {Nollert}(1996)}]{Nollert:1996rf}%
  \BibitemOpen
  \bibfield  {author} {\bibinfo {author} {\bibfnamefont {H.-P.}\ \bibnamefont {Nollert}},\ }\href {\doibase 10.1103/PhysRevD.53.4397} {\bibfield  {journal} {\bibinfo  {journal} {Phys. Rev. D}\ }\textbf {\bibinfo {volume} {53}},\ \bibinfo {pages} {4397} (\bibinfo {year} {1996})},\ \Eprint {http://arxiv.org/abs/gr-qc/9602032} {arXiv:gr-qc/9602032} \BibitemShut {NoStop}%
\bibitem [{\citenamefont {Daghigh}\ \emph {et~al.}(2020)\citenamefont {Daghigh}, \citenamefont {Green},\ and\ \citenamefont {Morey}}]{Daghigh:2020jyk}%
  \BibitemOpen
  \bibfield  {author} {\bibinfo {author} {\bibfnamefont {R.~G.}\ \bibnamefont {Daghigh}}, \bibinfo {author} {\bibfnamefont {M.~D.}\ \bibnamefont {Green}}, \ and\ \bibinfo {author} {\bibfnamefont {J.~C.}\ \bibnamefont {Morey}},\ }\href {\doibase 10.1103/PhysRevD.101.104009} {\bibfield  {journal} {\bibinfo  {journal} {Phys. Rev. D}\ }\textbf {\bibinfo {volume} {101}},\ \bibinfo {pages} {104009} (\bibinfo {year} {2020})},\ \Eprint {http://arxiv.org/abs/2002.07251} {arXiv:2002.07251 [gr-qc]} \BibitemShut {NoStop}%
\bibitem [{\citenamefont {Leung}\ \emph {et~al.}(1997)\citenamefont {Leung}, \citenamefont {Liu}, \citenamefont {Suen}, \citenamefont {Tam},\ and\ \citenamefont {Young}}]{Leung:1997was}%
  \BibitemOpen
  \bibfield  {author} {\bibinfo {author} {\bibfnamefont {P.~T.}\ \bibnamefont {Leung}}, \bibinfo {author} {\bibfnamefont {Y.~T.}\ \bibnamefont {Liu}}, \bibinfo {author} {\bibfnamefont {W.~M.}\ \bibnamefont {Suen}}, \bibinfo {author} {\bibfnamefont {C.~Y.}\ \bibnamefont {Tam}}, \ and\ \bibinfo {author} {\bibfnamefont {K.}~\bibnamefont {Young}},\ }\href {\doibase 10.1103/PhysRevLett.78.2894} {\bibfield  {journal} {\bibinfo  {journal} {Phys. Rev. Lett.}\ }\textbf {\bibinfo {volume} {78}},\ \bibinfo {pages} {2894} (\bibinfo {year} {1997})},\ \Eprint {http://arxiv.org/abs/gr-qc/9903031} {arXiv:gr-qc/9903031} \BibitemShut {NoStop}%
\bibitem [{\citenamefont {Leung}\ \emph {et~al.}(1999)\citenamefont {Leung}, \citenamefont {Liu}, \citenamefont {Suen}, \citenamefont {Tam},\ and\ \citenamefont {Young}}]{Leung:1999iq}%
  \BibitemOpen
  \bibfield  {author} {\bibinfo {author} {\bibfnamefont {P.~T.}\ \bibnamefont {Leung}}, \bibinfo {author} {\bibfnamefont {Y.~T.}\ \bibnamefont {Liu}}, \bibinfo {author} {\bibfnamefont {W.~M.}\ \bibnamefont {Suen}}, \bibinfo {author} {\bibfnamefont {C.~Y.}\ \bibnamefont {Tam}}, \ and\ \bibinfo {author} {\bibfnamefont {K.}~\bibnamefont {Young}},\ }\href {\doibase 10.1103/PhysRevD.59.044034} {\bibfield  {journal} {\bibinfo  {journal} {Phys. Rev. D}\ }\textbf {\bibinfo {volume} {59}},\ \bibinfo {pages} {044034} (\bibinfo {year} {1999})},\ \Eprint {http://arxiv.org/abs/gr-qc/9903032} {arXiv:gr-qc/9903032} \BibitemShut {NoStop}%
\bibitem [{\citenamefont {Barausse}\ \emph {et~al.}(2015)\citenamefont {Barausse}, \citenamefont {Cardoso},\ and\ \citenamefont {Pani}}]{Barausse:2014pra}%
  \BibitemOpen
  \bibfield  {author} {\bibinfo {author} {\bibfnamefont {E.}~\bibnamefont {Barausse}}, \bibinfo {author} {\bibfnamefont {V.}~\bibnamefont {Cardoso}}, \ and\ \bibinfo {author} {\bibfnamefont {P.}~\bibnamefont {Pani}},\ }\href {\doibase 10.1088/1742-6596/610/1/012044} {\bibfield  {journal} {\bibinfo  {journal} {J. Phys. Conf. Ser.}\ }\textbf {\bibinfo {volume} {610}},\ \bibinfo {pages} {012044} (\bibinfo {year} {2015})},\ \Eprint {http://arxiv.org/abs/1404.7140} {arXiv:1404.7140 [astro-ph.CO]} \BibitemShut {NoStop}%
\bibitem [{\citenamefont {Cheung}\ \emph {et~al.}(2022)\citenamefont {Cheung}, \citenamefont {Destounis}, \citenamefont {Macedo}, \citenamefont {Berti},\ and\ \citenamefont {Cardoso}}]{Cheung:2021bol}%
  \BibitemOpen
  \bibfield  {author} {\bibinfo {author} {\bibfnamefont {M.~H.-Y.}\ \bibnamefont {Cheung}}, \bibinfo {author} {\bibfnamefont {K.}~\bibnamefont {Destounis}}, \bibinfo {author} {\bibfnamefont {R.~P.}\ \bibnamefont {Macedo}}, \bibinfo {author} {\bibfnamefont {E.}~\bibnamefont {Berti}}, \ and\ \bibinfo {author} {\bibfnamefont {V.}~\bibnamefont {Cardoso}},\ }\href {\doibase 10.1103/PhysRevLett.128.111103} {\bibfield  {journal} {\bibinfo  {journal} {Phys. Rev. Lett.}\ }\textbf {\bibinfo {volume} {128}},\ \bibinfo {pages} {111103} (\bibinfo {year} {2022})},\ \Eprint {http://arxiv.org/abs/2111.05415} {arXiv:2111.05415 [gr-qc]} \BibitemShut {NoStop}%
\bibitem [{\citenamefont {Berti}\ \emph {et~al.}(2022)\citenamefont {Berti}, \citenamefont {Cardoso}, \citenamefont {Cheung}, \citenamefont {Di~Filippo}, \citenamefont {Duque}, \citenamefont {Martens},\ and\ \citenamefont {Mukohyama}}]{Berti:2022xfj}%
  \BibitemOpen
  \bibfield  {author} {\bibinfo {author} {\bibfnamefont {E.}~\bibnamefont {Berti}}, \bibinfo {author} {\bibfnamefont {V.}~\bibnamefont {Cardoso}}, \bibinfo {author} {\bibfnamefont {M.~H.-Y.}\ \bibnamefont {Cheung}}, \bibinfo {author} {\bibfnamefont {F.}~\bibnamefont {Di~Filippo}}, \bibinfo {author} {\bibfnamefont {F.}~\bibnamefont {Duque}}, \bibinfo {author} {\bibfnamefont {P.}~\bibnamefont {Martens}}, \ and\ \bibinfo {author} {\bibfnamefont {S.}~\bibnamefont {Mukohyama}},\ }\href {\doibase 10.1103/PhysRevD.106.084011} {\bibfield  {journal} {\bibinfo  {journal} {Phys. Rev. D}\ }\textbf {\bibinfo {volume} {106}},\ \bibinfo {pages} {084011} (\bibinfo {year} {2022})},\ \Eprint {http://arxiv.org/abs/2205.08547} {arXiv:2205.08547 [gr-qc]} \BibitemShut {NoStop}%
\bibitem [{\citenamefont {Cardoso}\ \emph {et~al.}(2016{\natexlab{a}})\citenamefont {Cardoso}, \citenamefont {Franzin},\ and\ \citenamefont {Pani}}]{Cardoso:2016rao}%
  \BibitemOpen
  \bibfield  {author} {\bibinfo {author} {\bibfnamefont {V.}~\bibnamefont {Cardoso}}, \bibinfo {author} {\bibfnamefont {E.}~\bibnamefont {Franzin}}, \ and\ \bibinfo {author} {\bibfnamefont {P.}~\bibnamefont {Pani}},\ }\href {\doibase 10.1103/PhysRevLett.116.171101} {\bibfield  {journal} {\bibinfo  {journal} {Phys. Rev. Lett.}\ }\textbf {\bibinfo {volume} {116}},\ \bibinfo {pages} {171101} (\bibinfo {year} {2016}{\natexlab{a}})},\ \bibinfo {note} {[Erratum: Phys.Rev.Lett. 117, 089902 (2016)]},\ \Eprint {http://arxiv.org/abs/1602.07309} {arXiv:1602.07309 [gr-qc]} \BibitemShut {NoStop}%
\bibitem [{\citenamefont {Cardoso}\ \emph {et~al.}(2016{\natexlab{b}})\citenamefont {Cardoso}, \citenamefont {Hopper}, \citenamefont {Macedo}, \citenamefont {Palenzuela},\ and\ \citenamefont {Pani}}]{Cardoso:2016oxy}%
  \BibitemOpen
  \bibfield  {author} {\bibinfo {author} {\bibfnamefont {V.}~\bibnamefont {Cardoso}}, \bibinfo {author} {\bibfnamefont {S.}~\bibnamefont {Hopper}}, \bibinfo {author} {\bibfnamefont {C.~F.~B.}\ \bibnamefont {Macedo}}, \bibinfo {author} {\bibfnamefont {C.}~\bibnamefont {Palenzuela}}, \ and\ \bibinfo {author} {\bibfnamefont {P.}~\bibnamefont {Pani}},\ }\href {\doibase 10.1103/PhysRevD.94.084031} {\bibfield  {journal} {\bibinfo  {journal} {Phys. Rev. D}\ }\textbf {\bibinfo {volume} {94}},\ \bibinfo {pages} {084031} (\bibinfo {year} {2016}{\natexlab{b}})},\ \Eprint {http://arxiv.org/abs/1608.08637} {arXiv:1608.08637 [gr-qc]} \BibitemShut {NoStop}%
\bibitem [{\citenamefont {Cardoso}\ and\ \citenamefont {Pani}(2017)}]{Cardoso:2017cqb}%
  \BibitemOpen
  \bibfield  {author} {\bibinfo {author} {\bibfnamefont {V.}~\bibnamefont {Cardoso}}\ and\ \bibinfo {author} {\bibfnamefont {P.}~\bibnamefont {Pani}},\ }\href {\doibase 10.1038/s41550-017-0225-y} {\bibfield  {journal} {\bibinfo  {journal} {Nature Astron.}\ }\textbf {\bibinfo {volume} {1}},\ \bibinfo {pages} {586} (\bibinfo {year} {2017})},\ \Eprint {http://arxiv.org/abs/1709.01525} {arXiv:1709.01525 [gr-qc]} \BibitemShut {NoStop}%
\bibitem [{\citenamefont {Abedi}\ \emph {et~al.}(2020)\citenamefont {Abedi}, \citenamefont {Afshordi}, \citenamefont {Oshita},\ and\ \citenamefont {Wang}}]{Abedi:2020ujo}%
  \BibitemOpen
  \bibfield  {author} {\bibinfo {author} {\bibfnamefont {J.}~\bibnamefont {Abedi}}, \bibinfo {author} {\bibfnamefont {N.}~\bibnamefont {Afshordi}}, \bibinfo {author} {\bibfnamefont {N.}~\bibnamefont {Oshita}}, \ and\ \bibinfo {author} {\bibfnamefont {Q.}~\bibnamefont {Wang}},\ }\href {\doibase 10.3390/universe6030043} {\bibfield  {journal} {\bibinfo  {journal} {Universe}\ }\textbf {\bibinfo {volume} {6}},\ \bibinfo {pages} {43} (\bibinfo {year} {2020})},\ \Eprint {http://arxiv.org/abs/2001.09553} {arXiv:2001.09553 [gr-qc]} \BibitemShut {NoStop}%
\bibitem [{\citenamefont {Jaramillo}\ \emph {et~al.}(2021)\citenamefont {Jaramillo}, \citenamefont {Panosso~Macedo},\ and\ \citenamefont {Al~Sheikh}}]{Jaramillo:2020tuu}%
  \BibitemOpen
  \bibfield  {author} {\bibinfo {author} {\bibfnamefont {J.~L.}\ \bibnamefont {Jaramillo}}, \bibinfo {author} {\bibfnamefont {R.}~\bibnamefont {Panosso~Macedo}}, \ and\ \bibinfo {author} {\bibfnamefont {L.}~\bibnamefont {Al~Sheikh}},\ }\href {\doibase 10.1103/PhysRevX.11.031003} {\bibfield  {journal} {\bibinfo  {journal} {Phys. Rev. X}\ }\textbf {\bibinfo {volume} {11}},\ \bibinfo {pages} {031003} (\bibinfo {year} {2021})},\ \Eprint {http://arxiv.org/abs/2004.06434} {arXiv:2004.06434 [gr-qc]} \BibitemShut {NoStop}%
\bibitem [{\citenamefont {Destounis}\ \emph {et~al.}(2021)\citenamefont {Destounis}, \citenamefont {Macedo}, \citenamefont {Berti}, \citenamefont {Cardoso},\ and\ \citenamefont {Jaramillo}}]{Destounis:2021lum}%
  \BibitemOpen
  \bibfield  {author} {\bibinfo {author} {\bibfnamefont {K.}~\bibnamefont {Destounis}}, \bibinfo {author} {\bibfnamefont {R.~P.}\ \bibnamefont {Macedo}}, \bibinfo {author} {\bibfnamefont {E.}~\bibnamefont {Berti}}, \bibinfo {author} {\bibfnamefont {V.}~\bibnamefont {Cardoso}}, \ and\ \bibinfo {author} {\bibfnamefont {J.~L.}\ \bibnamefont {Jaramillo}},\ }\href {\doibase 10.1103/PhysRevD.104.084091} {\bibfield  {journal} {\bibinfo  {journal} {Phys. Rev. D}\ }\textbf {\bibinfo {volume} {104}},\ \bibinfo {pages} {084091} (\bibinfo {year} {2021})},\ \Eprint {http://arxiv.org/abs/2107.09673} {arXiv:2107.09673 [gr-qc]} \BibitemShut {NoStop}%
\bibitem [{\citenamefont {Gasperin}\ and\ \citenamefont {Jaramillo}(2022)}]{Gasperin:2021kfv}%
  \BibitemOpen
  \bibfield  {author} {\bibinfo {author} {\bibfnamefont {E.}~\bibnamefont {Gasperin}}\ and\ \bibinfo {author} {\bibfnamefont {J.~L.}\ \bibnamefont {Jaramillo}},\ }\href {\doibase 10.1088/1361-6382/ac5054} {\bibfield  {journal} {\bibinfo  {journal} {Class. Quant. Grav.}\ }\textbf {\bibinfo {volume} {39}},\ \bibinfo {pages} {115010} (\bibinfo {year} {2022})},\ \Eprint {http://arxiv.org/abs/2107.12865} {arXiv:2107.12865 [gr-qc]} \BibitemShut {NoStop}%
\bibitem [{\citenamefont {Boyanov}\ \emph {et~al.}(2023)\citenamefont {Boyanov}, \citenamefont {Destounis}, \citenamefont {Panosso~Macedo}, \citenamefont {Cardoso},\ and\ \citenamefont {Jaramillo}}]{Boyanov:2022ark}%
  \BibitemOpen
  \bibfield  {author} {\bibinfo {author} {\bibfnamefont {V.}~\bibnamefont {Boyanov}}, \bibinfo {author} {\bibfnamefont {K.}~\bibnamefont {Destounis}}, \bibinfo {author} {\bibfnamefont {R.}~\bibnamefont {Panosso~Macedo}}, \bibinfo {author} {\bibfnamefont {V.}~\bibnamefont {Cardoso}}, \ and\ \bibinfo {author} {\bibfnamefont {J.~L.}\ \bibnamefont {Jaramillo}},\ }\href {\doibase 10.1103/PhysRevD.107.064012} {\bibfield  {journal} {\bibinfo  {journal} {Phys. Rev. D}\ }\textbf {\bibinfo {volume} {107}},\ \bibinfo {pages} {064012} (\bibinfo {year} {2023})},\ \Eprint {http://arxiv.org/abs/2209.12950} {arXiv:2209.12950 [gr-qc]} \BibitemShut {NoStop}%
\bibitem [{\citenamefont {Jaramillo}(2022)}]{Jaramillo:2022kuv}%
  \BibitemOpen
  \bibfield  {author} {\bibinfo {author} {\bibfnamefont {J.~L.}\ \bibnamefont {Jaramillo}},\ }\href {\doibase 10.1088/1361-6382/ac8ddc} {\bibfield  {journal} {\bibinfo  {journal} {Class. Quant. Grav.}\ }\textbf {\bibinfo {volume} {39}},\ \bibinfo {pages} {217002} (\bibinfo {year} {2022})},\ \Eprint {http://arxiv.org/abs/2206.08025} {arXiv:2206.08025 [gr-qc]} \BibitemShut {NoStop}%
\bibitem [{\citenamefont {Kyutoku}\ \emph {et~al.}(2023)\citenamefont {Kyutoku}, \citenamefont {Motohashi},\ and\ \citenamefont {Tanaka}}]{Kyutoku:2022gbr}%
  \BibitemOpen
  \bibfield  {author} {\bibinfo {author} {\bibfnamefont {K.}~\bibnamefont {Kyutoku}}, \bibinfo {author} {\bibfnamefont {H.}~\bibnamefont {Motohashi}}, \ and\ \bibinfo {author} {\bibfnamefont {T.}~\bibnamefont {Tanaka}},\ }\href {\doibase 10.1103/PhysRevD.107.044012} {\bibfield  {journal} {\bibinfo  {journal} {Phys. Rev. D}\ }\textbf {\bibinfo {volume} {107}},\ \bibinfo {pages} {044012} (\bibinfo {year} {2023})},\ \Eprint {http://arxiv.org/abs/2206.00671} {arXiv:2206.00671 [gr-qc]} \BibitemShut {NoStop}%
\bibitem [{\citenamefont {Sarkar}\ \emph {et~al.}(2023)\citenamefont {Sarkar}, \citenamefont {Rahman},\ and\ \citenamefont {Chakraborty}}]{Sarkar:2023rhp}%
  \BibitemOpen
  \bibfield  {author} {\bibinfo {author} {\bibfnamefont {S.}~\bibnamefont {Sarkar}}, \bibinfo {author} {\bibfnamefont {M.}~\bibnamefont {Rahman}}, \ and\ \bibinfo {author} {\bibfnamefont {S.}~\bibnamefont {Chakraborty}},\ }\href {\doibase 10.1103/PhysRevD.108.104002} {\bibfield  {journal} {\bibinfo  {journal} {Phys. Rev. D}\ }\textbf {\bibinfo {volume} {108}},\ \bibinfo {pages} {104002} (\bibinfo {year} {2023})},\ \Eprint {http://arxiv.org/abs/2304.06829} {arXiv:2304.06829 [gr-qc]} \BibitemShut {NoStop}%
\bibitem [{\citenamefont {Destounis}\ \emph {et~al.}(2024)\citenamefont {Destounis}, \citenamefont {Boyanov},\ and\ \citenamefont {Panosso~Macedo}}]{Destounis:2023nmb}%
  \BibitemOpen
  \bibfield  {author} {\bibinfo {author} {\bibfnamefont {K.}~\bibnamefont {Destounis}}, \bibinfo {author} {\bibfnamefont {V.}~\bibnamefont {Boyanov}}, \ and\ \bibinfo {author} {\bibfnamefont {R.}~\bibnamefont {Panosso~Macedo}},\ }\href {\doibase 10.1103/PhysRevD.109.044023} {\bibfield  {journal} {\bibinfo  {journal} {Phys. Rev. D}\ }\textbf {\bibinfo {volume} {109}},\ \bibinfo {pages} {044023} (\bibinfo {year} {2024})},\ \Eprint {http://arxiv.org/abs/2312.11630} {arXiv:2312.11630 [gr-qc]} \BibitemShut {NoStop}%
\bibitem [{\citenamefont {Are\'an}\ \emph {et~al.}(2023)\citenamefont {Are\'an}, \citenamefont {Fari\~na},\ and\ \citenamefont {Landsteiner}}]{Arean:2023ejh}%
  \BibitemOpen
  \bibfield  {author} {\bibinfo {author} {\bibfnamefont {D.}~\bibnamefont {Are\'an}}, \bibinfo {author} {\bibfnamefont {D.~G.}\ \bibnamefont {Fari\~na}}, \ and\ \bibinfo {author} {\bibfnamefont {K.}~\bibnamefont {Landsteiner}},\ }\href {\doibase 10.1007/JHEP12(2023)187} {\bibfield  {journal} {\bibinfo  {journal} {JHEP}\ }\textbf {\bibinfo {volume} {12}},\ \bibinfo {pages} {187} (\bibinfo {year} {2023})},\ \Eprint {http://arxiv.org/abs/2307.08751} {arXiv:2307.08751 [hep-th]} \BibitemShut {NoStop}%
\bibitem [{\citenamefont {Cownden}\ \emph {et~al.}(2024)\citenamefont {Cownden}, \citenamefont {Pantelidou},\ and\ \citenamefont {Zilh\~ao}}]{Cownden:2023dam}%
  \BibitemOpen
  \bibfield  {author} {\bibinfo {author} {\bibfnamefont {B.}~\bibnamefont {Cownden}}, \bibinfo {author} {\bibfnamefont {C.}~\bibnamefont {Pantelidou}}, \ and\ \bibinfo {author} {\bibfnamefont {M.}~\bibnamefont {Zilh\~ao}},\ }\href {\doibase 10.1007/JHEP05(2024)202} {\bibfield  {journal} {\bibinfo  {journal} {JHEP}\ }\textbf {\bibinfo {volume} {05}},\ \bibinfo {pages} {202} (\bibinfo {year} {2024})},\ \Eprint {http://arxiv.org/abs/2312.08352} {arXiv:2312.08352 [gr-qc]} \BibitemShut {NoStop}%
\bibitem [{\citenamefont {Destounis}\ and\ \citenamefont {Duque}(2023)}]{Destounis:2023ruj}%
  \BibitemOpen
  \bibfield  {author} {\bibinfo {author} {\bibfnamefont {K.}~\bibnamefont {Destounis}}\ and\ \bibinfo {author} {\bibfnamefont {F.}~\bibnamefont {Duque}}\ }(\bibinfo {year} {2023})\ \Eprint {http://arxiv.org/abs/2308.16227} {arXiv:2308.16227 [gr-qc]} \BibitemShut {NoStop}%
\bibitem [{\citenamefont {Courty}\ \emph {et~al.}(2023)\citenamefont {Courty}, \citenamefont {Destounis},\ and\ \citenamefont {Pani}}]{Courty:2023rxk}%
  \BibitemOpen
  \bibfield  {author} {\bibinfo {author} {\bibfnamefont {A.}~\bibnamefont {Courty}}, \bibinfo {author} {\bibfnamefont {K.}~\bibnamefont {Destounis}}, \ and\ \bibinfo {author} {\bibfnamefont {P.}~\bibnamefont {Pani}},\ }\href {\doibase 10.1103/PhysRevD.108.104027} {\bibfield  {journal} {\bibinfo  {journal} {Phys. Rev. D}\ }\textbf {\bibinfo {volume} {108}},\ \bibinfo {pages} {104027} (\bibinfo {year} {2023})},\ \Eprint {http://arxiv.org/abs/2307.11155} {arXiv:2307.11155 [gr-qc]} \BibitemShut {NoStop}%
\bibitem [{\citenamefont {Boyanov}\ \emph {et~al.}(2024)\citenamefont {Boyanov}, \citenamefont {Cardoso}, \citenamefont {Destounis}, \citenamefont {Jaramillo},\ and\ \citenamefont {Panosso~Macedo}}]{Boyanov:2023qqf}%
  \BibitemOpen
  \bibfield  {author} {\bibinfo {author} {\bibfnamefont {V.}~\bibnamefont {Boyanov}}, \bibinfo {author} {\bibfnamefont {V.}~\bibnamefont {Cardoso}}, \bibinfo {author} {\bibfnamefont {K.}~\bibnamefont {Destounis}}, \bibinfo {author} {\bibfnamefont {J.~L.}\ \bibnamefont {Jaramillo}}, \ and\ \bibinfo {author} {\bibfnamefont {R.}~\bibnamefont {Panosso~Macedo}},\ }\href {\doibase 10.1103/PhysRevD.109.064068} {\bibfield  {journal} {\bibinfo  {journal} {Phys. Rev. D}\ }\textbf {\bibinfo {volume} {109}},\ \bibinfo {pages} {064068} (\bibinfo {year} {2024})},\ \Eprint {http://arxiv.org/abs/2312.11998} {arXiv:2312.11998 [gr-qc]} \BibitemShut {NoStop}%
\bibitem [{\citenamefont {Cao}\ \emph {et~al.}(2024)\citenamefont {Cao}, \citenamefont {Chen}, \citenamefont {Wu}, \citenamefont {Xie},\ and\ \citenamefont {Zhou}}]{Cao:2024oud}%
  \BibitemOpen
  \bibfield  {author} {\bibinfo {author} {\bibfnamefont {L.-M.}\ \bibnamefont {Cao}}, \bibinfo {author} {\bibfnamefont {J.-N.}\ \bibnamefont {Chen}}, \bibinfo {author} {\bibfnamefont {L.-B.}\ \bibnamefont {Wu}}, \bibinfo {author} {\bibfnamefont {L.}~\bibnamefont {Xie}}, \ and\ \bibinfo {author} {\bibfnamefont {Y.-S.}\ \bibnamefont {Zhou}},\ }\href@noop {} {\  (\bibinfo {year} {2024})},\ \Eprint {http://arxiv.org/abs/2401.09907} {arXiv:2401.09907 [gr-qc]} \BibitemShut {NoStop}%
\bibitem [{\citenamefont {Cardoso}\ \emph {et~al.}(2024)\citenamefont {Cardoso}, \citenamefont {Kastha},\ and\ \citenamefont {Panosso~Macedo}}]{Cardoso:2024mrw}%
  \BibitemOpen
  \bibfield  {author} {\bibinfo {author} {\bibfnamefont {V.}~\bibnamefont {Cardoso}}, \bibinfo {author} {\bibfnamefont {S.}~\bibnamefont {Kastha}}, \ and\ \bibinfo {author} {\bibfnamefont {R.}~\bibnamefont {Panosso~Macedo}},\ }\href@noop {} {\  (\bibinfo {year} {2024})},\ \Eprint {http://arxiv.org/abs/2404.01374} {arXiv:2404.01374 [gr-qc]} \BibitemShut {NoStop}%
\bibitem [{\citenamefont {Rosato}\ \emph {et~al.}(2024)\citenamefont {Rosato}, \citenamefont {Destounis},\ and\ \citenamefont {Pani}}]{Rosato:2024arw}%
  \BibitemOpen
  \bibfield  {author} {\bibinfo {author} {\bibfnamefont {R.~F.}\ \bibnamefont {Rosato}}, \bibinfo {author} {\bibfnamefont {K.}~\bibnamefont {Destounis}}, \ and\ \bibinfo {author} {\bibfnamefont {P.}~\bibnamefont {Pani}},\ }\href@noop {} {\  (\bibinfo {year} {2024})},\ \Eprint {http://arxiv.org/abs/2406.01692} {arXiv:2406.01692 [gr-qc]} \BibitemShut {NoStop}%
\bibitem [{\citenamefont {Bena}\ \emph {et~al.}(2022)\citenamefont {Bena}, \citenamefont {Martinec}, \citenamefont {Mathur},\ and\ \citenamefont {Warner}}]{Bena:2022rna}%
  \BibitemOpen
  \bibfield  {author} {\bibinfo {author} {\bibfnamefont {I.}~\bibnamefont {Bena}}, \bibinfo {author} {\bibfnamefont {E.~J.}\ \bibnamefont {Martinec}}, \bibinfo {author} {\bibfnamefont {S.~D.}\ \bibnamefont {Mathur}}, \ and\ \bibinfo {author} {\bibfnamefont {N.~P.}\ \bibnamefont {Warner}},\ }\href@noop {} {\  (\bibinfo {year} {2022})},\ \Eprint {http://arxiv.org/abs/2204.13113} {arXiv:2204.13113 [hep-th]} \BibitemShut {NoStop}%
\bibitem [{\citenamefont {Almheiri}\ \emph {et~al.}(2013)\citenamefont {Almheiri}, \citenamefont {Marolf}, \citenamefont {Polchinski},\ and\ \citenamefont {Sully}}]{Almheiri:2012rt}%
  \BibitemOpen
  \bibfield  {author} {\bibinfo {author} {\bibfnamefont {A.}~\bibnamefont {Almheiri}}, \bibinfo {author} {\bibfnamefont {D.}~\bibnamefont {Marolf}}, \bibinfo {author} {\bibfnamefont {J.}~\bibnamefont {Polchinski}}, \ and\ \bibinfo {author} {\bibfnamefont {J.}~\bibnamefont {Sully}},\ }\href {\doibase 10.1007/JHEP02(2013)062} {\bibfield  {journal} {\bibinfo  {journal} {JHEP}\ }\textbf {\bibinfo {volume} {02}},\ \bibinfo {pages} {062} (\bibinfo {year} {2013})},\ \Eprint {http://arxiv.org/abs/1207.3123} {arXiv:1207.3123 [hep-th]} \BibitemShut {NoStop}%
\bibitem [{\citenamefont {Grosso}\ and\ \citenamefont {Parravicini}(2000)}]{GROSSO20001}%
  \BibitemOpen
  \bibfield  {author} {\bibinfo {author} {\bibfnamefont {G.}~\bibnamefont {Grosso}}\ and\ \bibinfo {author} {\bibfnamefont {G.~P.}\ \bibnamefont {Parravicini}},\ }in\ \href {\doibase https://doi.org/10.1016/B978-012304460-0/50001-3} {\emph {\bibinfo {booktitle} {Solid State Physics}}},\ \bibinfo {editor} {edited by\ \bibinfo {editor} {\bibfnamefont {G.}~\bibnamefont {Grosso}}\ and\ \bibinfo {editor} {\bibfnamefont {G.~P.}\ \bibnamefont {Parravicini}}}\ (\bibinfo  {publisher} {Academic Press},\ \bibinfo {address} {London},\ \bibinfo {year} {2000})\ pp.\ \bibinfo {pages} {1--36}\BibitemShut {NoStop}%
\bibitem [{\citenamefont {Mirbabayi}(2020)}]{Mirbabayi:2018mdm}%
  \BibitemOpen
  \bibfield  {author} {\bibinfo {author} {\bibfnamefont {M.}~\bibnamefont {Mirbabayi}},\ }\href {\doibase 10.1088/1475-7516/2020/01/052} {\bibfield  {journal} {\bibinfo  {journal} {JCAP}\ }\textbf {\bibinfo {volume} {01}},\ \bibinfo {pages} {052} (\bibinfo {year} {2020})},\ \Eprint {http://arxiv.org/abs/1807.04843} {arXiv:1807.04843 [gr-qc]} \BibitemShut {NoStop}%
\bibitem [{\citenamefont {Hui}\ \emph {et~al.}(2019)\citenamefont {Hui}, \citenamefont {Kabat},\ and\ \citenamefont {Wong}}]{Hui:2019aox}%
  \BibitemOpen
  \bibfield  {author} {\bibinfo {author} {\bibfnamefont {L.}~\bibnamefont {Hui}}, \bibinfo {author} {\bibfnamefont {D.}~\bibnamefont {Kabat}}, \ and\ \bibinfo {author} {\bibfnamefont {S.~S.~C.}\ \bibnamefont {Wong}},\ }\href {\doibase 10.1088/1475-7516/2019/12/020} {\bibfield  {journal} {\bibinfo  {journal} {JCAP}\ }\textbf {\bibinfo {volume} {12}},\ \bibinfo {pages} {020} (\bibinfo {year} {2019})},\ \Eprint {http://arxiv.org/abs/1909.10382} {arXiv:1909.10382 [gr-qc]} \BibitemShut {NoStop}%
\bibitem [{\citenamefont {Harmark}\ \emph {et~al.}(2010)\citenamefont {Harmark}, \citenamefont {Natario},\ and\ \citenamefont {Schiappa}}]{Harmark:2007jy}%
  \BibitemOpen
  \bibfield  {author} {\bibinfo {author} {\bibfnamefont {T.}~\bibnamefont {Harmark}}, \bibinfo {author} {\bibfnamefont {J.}~\bibnamefont {Natario}}, \ and\ \bibinfo {author} {\bibfnamefont {R.}~\bibnamefont {Schiappa}},\ }\href {\doibase 10.4310/ATMP.2010.v14.n3.a1} {\bibfield  {journal} {\bibinfo  {journal} {Adv. Theor. Math. Phys.}\ }\textbf {\bibinfo {volume} {14}},\ \bibinfo {pages} {727} (\bibinfo {year} {2010})},\ \Eprint {http://arxiv.org/abs/0708.0017} {arXiv:0708.0017 [hep-th]} \BibitemShut {NoStop}%
\bibitem [{\citenamefont {Oshita}\ \emph {et~al.}(2024)\citenamefont {Oshita}, \citenamefont {Takahashi},\ and\ \citenamefont {Mukohyama}}]{Oshita:2024fzf}%
  \BibitemOpen
  \bibfield  {author} {\bibinfo {author} {\bibfnamefont {N.}~\bibnamefont {Oshita}}, \bibinfo {author} {\bibfnamefont {K.}~\bibnamefont {Takahashi}}, \ and\ \bibinfo {author} {\bibfnamefont {S.}~\bibnamefont {Mukohyama}},\ }\href@noop {} {\  (\bibinfo {year} {2024})},\ \Eprint {http://arxiv.org/abs/2406.04525} {arXiv:2406.04525 [gr-qc]} \BibitemShut {NoStop}%
\bibitem [{\citenamefont {{Goebel}}(1972)}]{Goebel}%
  \BibitemOpen
  \bibfield  {author} {\bibinfo {author} {\bibfnamefont {C.~J.}\ \bibnamefont {{Goebel}}},\ }\href {\doibase 10.1086/180898} {\bibfield  {journal} {\bibinfo  {journal} {apjl}\ }\textbf {\bibinfo {volume} {172}},\ \bibinfo {pages} {L95} (\bibinfo {year} {1972})}\BibitemShut {NoStop}%
\bibitem [{\citenamefont {Ferrari}\ and\ \citenamefont {Mashhoon}(1984)}]{Ferrari:1984zz}%
  \BibitemOpen
  \bibfield  {author} {\bibinfo {author} {\bibfnamefont {V.}~\bibnamefont {Ferrari}}\ and\ \bibinfo {author} {\bibfnamefont {B.}~\bibnamefont {Mashhoon}},\ }\href {\doibase 10.1103/PhysRevD.30.295} {\bibfield  {journal} {\bibinfo  {journal} {Phys. Rev. D}\ }\textbf {\bibinfo {volume} {30}},\ \bibinfo {pages} {295} (\bibinfo {year} {1984})}\BibitemShut {NoStop}%
\bibitem [{\citenamefont {Mashhoon}(1985)}]{Mashhoon:1985cya}%
  \BibitemOpen
  \bibfield  {author} {\bibinfo {author} {\bibfnamefont {B.}~\bibnamefont {Mashhoon}},\ }\href {\doibase 10.1103/PhysRevD.31.290} {\bibfield  {journal} {\bibinfo  {journal} {Phys. Rev. D}\ }\textbf {\bibinfo {volume} {31}},\ \bibinfo {pages} {290} (\bibinfo {year} {1985})}\BibitemShut {NoStop}%
\bibitem [{\citenamefont {Berti}\ and\ \citenamefont {Kokkotas}(2005)}]{Berti:2005eb}%
  \BibitemOpen
  \bibfield  {author} {\bibinfo {author} {\bibfnamefont {E.}~\bibnamefont {Berti}}\ and\ \bibinfo {author} {\bibfnamefont {K.~D.}\ \bibnamefont {Kokkotas}},\ }\href {\doibase 10.1103/PhysRevD.71.124008} {\bibfield  {journal} {\bibinfo  {journal} {Phys. Rev. D}\ }\textbf {\bibinfo {volume} {71}},\ \bibinfo {pages} {124008} (\bibinfo {year} {2005})},\ \Eprint {http://arxiv.org/abs/gr-qc/0502065} {arXiv:gr-qc/0502065} \BibitemShut {NoStop}%
\bibitem [{\citenamefont {Cornish}\ and\ \citenamefont {Levin}(2003)}]{Cornish:2003ig}%
  \BibitemOpen
  \bibfield  {author} {\bibinfo {author} {\bibfnamefont {N.~J.}\ \bibnamefont {Cornish}}\ and\ \bibinfo {author} {\bibfnamefont {J.~J.}\ \bibnamefont {Levin}},\ }\href {\doibase 10.1088/0264-9381/20/9/304} {\bibfield  {journal} {\bibinfo  {journal} {Class. Quant. Grav.}\ }\textbf {\bibinfo {volume} {20}},\ \bibinfo {pages} {1649} (\bibinfo {year} {2003})},\ \Eprint {http://arxiv.org/abs/gr-qc/0304056} {arXiv:gr-qc/0304056} \BibitemShut {NoStop}%
\bibitem [{\citenamefont {Cardoso}\ \emph {et~al.}(2009)\citenamefont {Cardoso}, \citenamefont {Miranda}, \citenamefont {Berti}, \citenamefont {Witek},\ and\ \citenamefont {Zanchin}}]{Cardoso:2008bp}%
  \BibitemOpen
  \bibfield  {author} {\bibinfo {author} {\bibfnamefont {V.}~\bibnamefont {Cardoso}}, \bibinfo {author} {\bibfnamefont {A.~S.}\ \bibnamefont {Miranda}}, \bibinfo {author} {\bibfnamefont {E.}~\bibnamefont {Berti}}, \bibinfo {author} {\bibfnamefont {H.}~\bibnamefont {Witek}}, \ and\ \bibinfo {author} {\bibfnamefont {V.~T.}\ \bibnamefont {Zanchin}},\ }\href {\doibase 10.1103/PhysRevD.79.064016} {\bibfield  {journal} {\bibinfo  {journal} {Phys. Rev. D}\ }\textbf {\bibinfo {volume} {79}},\ \bibinfo {pages} {064016} (\bibinfo {year} {2009})},\ \Eprint {http://arxiv.org/abs/0812.1806} {arXiv:0812.1806 [hep-th]} \BibitemShut {NoStop}%
\bibitem [{\citenamefont {Ianniccari}\ \emph {et~al.}(2024)\citenamefont {Ianniccari}, \citenamefont {Iovino}, \citenamefont {Kehagias}, \citenamefont {Perrone},\ and\ \citenamefont {Riotto}}]{Ianniccari:2024ltb}%
  \BibitemOpen
  \bibfield  {author} {\bibinfo {author} {\bibfnamefont {A.}~\bibnamefont {Ianniccari}}, \bibinfo {author} {\bibfnamefont {A.~J.}\ \bibnamefont {Iovino}}, \bibinfo {author} {\bibfnamefont {A.}~\bibnamefont {Kehagias}}, \bibinfo {author} {\bibfnamefont {D.}~\bibnamefont {Perrone}}, \ and\ \bibinfo {author} {\bibfnamefont {A.}~\bibnamefont {Riotto}},\ }\href@noop {} {\  (\bibinfo {year} {2024})},\ \Eprint {http://arxiv.org/abs/2404.02801} {arXiv:2404.02801 [astro-ph.CO]} \BibitemShut {NoStop}%
\bibitem [{\citenamefont {Yang}\ \emph {et~al.}(2024)\citenamefont {Yang}, \citenamefont {Mai}, \citenamefont {Yang}, \citenamefont {Shao},\ and\ \citenamefont {Berti}}]{Yang}%
  \BibitemOpen
  \bibfield  {author} {\bibinfo {author} {\bibfnamefont {Y.}~\bibnamefont {Yang}}, \bibinfo {author} {\bibfnamefont {Z.-F.}\ \bibnamefont {Mai}}, \bibinfo {author} {\bibfnamefont {R.-Q.}\ \bibnamefont {Yang}}, \bibinfo {author} {\bibfnamefont {L.}~\bibnamefont {Shao}}, \ and\ \bibinfo {author} {\bibfnamefont {E.}~\bibnamefont {Berti}},\ }\href@noop {} {\emph {\bibinfo {title} {{Spectral instability of black holes: relating the frequency domain to the time domain}}}}\ (\bibinfo {year} {2024})\BibitemShut {NoStop}%
\bibitem [{\citenamefont {Fl{\"u}gge}(2012)}]{flugge2012practical}%
  \BibitemOpen
  \bibfield  {author} {\bibinfo {author} {\bibfnamefont {S.}~\bibnamefont {Fl{\"u}gge}},\ }\href@noop {} {\emph {\bibinfo {title} {Practical quantum mechanics}}}\ (\bibinfo  {publisher} {Springer Science \& Business Media},\ \bibinfo {year} {2012})\BibitemShut {NoStop}%
\bibitem [{\citenamefont {Iyer}\ and\ \citenamefont {Will}(1987)}]{transfer_will_wkb_PhysRevD.35.3621}%
  \BibitemOpen
  \bibfield  {author} {\bibinfo {author} {\bibfnamefont {S.}~\bibnamefont {Iyer}}\ and\ \bibinfo {author} {\bibfnamefont {C.~M.}\ \bibnamefont {Will}},\ }\href {\doibase 10.1103/PhysRevD.35.3621} {\bibfield  {journal} {\bibinfo  {journal} {Phys. Rev. D}\ }\textbf {\bibinfo {volume} {35}},\ \bibinfo {pages} {3621} (\bibinfo {year} {1987})}\BibitemShut {NoStop}%
\bibitem [{\citenamefont {Schutz}\ and\ \citenamefont {Will}(1985)}]{Schutz:1985km}%
  \BibitemOpen
  \bibfield  {author} {\bibinfo {author} {\bibfnamefont {B.~F.}\ \bibnamefont {Schutz}}\ and\ \bibinfo {author} {\bibfnamefont {C.~M.}\ \bibnamefont {Will}},\ }\href {\doibase 10.1086/184453} {\bibfield  {journal} {\bibinfo  {journal} {Astrophys. J. Lett.}\ }\textbf {\bibinfo {volume} {291}},\ \bibinfo {pages} {L33} (\bibinfo {year} {1985})}\BibitemShut {NoStop}%
\end{thebibliography}%

\clearpage
\newpage
\maketitle
\onecolumngrid
\begin{center}
\vspace{0.05in}
 
\vspace{0.05in}
{ \large\it Supplementary Material}
\end{center}
\onecolumngrid
\setcounter{equation}{0}
\setcounter{figure}{0}
\setcounter{section}{0}
\setcounter{table}{0}
\setcounter{page}{1}
\makeatletter
\renewcommand{\theequation}{S\arabic{equation}}
\renewcommand{\thefigure}{S\arabic{figure}}
\renewcommand{\thetable}{S\arabic{table}}

\section*{The transfer formalism}
\noindent
Multiple equivalent definitions for the transfer matrix can be used to describe the problem we are studying. In this section we make explicit our definitions and the relations we use to find the matrix form in Eq.~(\ref{eq:transferm}). 

We start from the definition of the 
free solution in the asymptotic region at the left and at the right of the localised potential as
\begin{equation}
    A_L e^{i\omega r_*} + B_L e^{-i\omega r_*},\quad r_*\to -\infty,
\end{equation}
\begin{equation}
    A_R e^{i\omega x} + B_R e^{-i\omega r_*},\quad r_*\to +\infty.
\end{equation}
We can write the two solutions as vectors by defining, as in Eq.~(\ref{eq:basis})
\begin{equation}
   e^{-i\omega r_*} \to \begin{pmatrix} 0\\1\end{pmatrix}, \quad e^{i\omega r_*} \to \begin{pmatrix} 1\\0\end{pmatrix},
\end{equation}
which allows us to write the problem with matrices.
We define the scattering matrix as
\begin{equation}
    S = \begin{pmatrix} r&  t'\\ t&  r'\end{pmatrix},
\end{equation}
where $r$, $t$ are transmission and reflection coefficients for a wave coming from the left and $r'$, $t'$ are the coefficients for a wave coming from the right, as in Fig.~\ref{fig:V1}.

The $S$ matrix connects the vectors
\begin{equation}
     S \begin{pmatrix}
       A_L\\ B_R  
   \end{pmatrix} =\begin{pmatrix}
       B_L\\ A_R  
   \end{pmatrix}, 
\end{equation}
such that we can define the standard scattering problem of a wave coming from the left and scattering on a potential as
\begin{equation}
   \begin{pmatrix} r&  t'\\ t&  r'\end{pmatrix} \begin{pmatrix}
       1\\0
   \end{pmatrix}  = \begin{pmatrix}
       r\\t
   \end{pmatrix},
\end{equation}
which gives the correct reflection and transmission coefficients for the solutions at the left and the right of the potential
\begin{equation}
    e^{i\omega r_*} + r\,e^{-i\omega r_*} \longleftrightarrow t\,e^{i \omega r_*},
\end{equation}
where the arrow represents the connection of the asymptotic states through the potential.

Having defined the scattering matrix in this way it is immediate to connect the reflection and transmission coefficients with the matrix elements of the transfer matrix
\begin{equation}
    M \begin{pmatrix}
       A_L\\ B_L 
   \end{pmatrix} =\begin{pmatrix}
       A_R\\ B_R  
   \end{pmatrix}. 
\end{equation}
This leads to the definition of the transfer matrix as
\begin{equation}
    M = \begin{pmatrix} t-r r'/t'&  r'/t'\\ -r/t'&  1/t'\end{pmatrix},
\end{equation}
as in Eq.~(\ref{eq:transferm}).\\
Furthermore we can use the property $S S^{\dagger}=1$ and $S^{\dagger}S=1$ to write
\begin{equation}
    |r|^2 + |t'|^2=1, \quad |r'|^2 + |t|^2=1, \quad r t^* =- t' r'^*
\end{equation}
\begin{equation}
    |r|^2 + |t|^2=1, \quad |r'|^2 + |t'|^2=1, \quad r^* t' =- t^* r'.
\end{equation}
If the problem is symmetric under time reversal there is also the additional relation $t=t'$,
which implies
\begin{equation}
    r=-r'^*\frac{t}{t^*}.
\end{equation}

\section{Coefficients for the P\"oschl-Teller  and Schwartzschild potentials}
\noindent
For the P\"oschl-Teller potential with
\begin{equation}
    V_1=\frac{\lambda \,(1-\lambda)}{r_s^2}{\rm sech}^2\left(\frac{r_*-c}{r_s}\right), 
\end{equation}
the transfer matrix is (see for instance Ref.~\cite{flugge2012practical})
\begin{equation}
    M= 
    \begin{pmatrix}
        \frac{\Gamma(i\omega r_s +1)\Gamma(i\omega r_s )}{\Gamma(i\omega r_s  +1 - \lambda)\Gamma(i\omega r_s  + \lambda)} & \frac{\Gamma(1-i\omega r_s )\Gamma(i\omega r_s )}{\Gamma(1-\lambda)\Gamma(\lambda)}\\
        & \\
       \frac{\Gamma(1+i\omega r_s )\Gamma(-i\omega r_s )}{\Gamma(1-\lambda)\Gamma\left(\lambda\right)}& \frac{\Gamma\left(1-i\omega r_s  \right)\Gamma\left(-i\omega r_s \right)}{\Gamma\left(1-i\omega r_s - \lambda\right)\Gamma\left(\lambda-i \omega r_s \right)},
    \end{pmatrix}
\end{equation}
and therefore 
\begin{equation}
\label{eq:pt_ref}
    r'= \frac{\Gamma(i\omega r_s)\Gamma(1-i\omega r_s  - \lambda)\Gamma(\lambda-i\omega r_s)}{\Gamma(1-\lambda )\Gamma(\lambda)\Gamma(-i\omega r_s)}, \quad r= -\frac{\Gamma(1+i\omega r_s)\Gamma(1-i\omega r_s  - \lambda)\Gamma(\lambda-i\omega r_s)}{\Gamma(1-\lambda )\Gamma(\lambda)\Gamma(1-i\omega r_s )}.
\end{equation}
For the potential in Eq. (\ref{eq:potentials}) we have  $\lambda=\epsilon \to 0$ and  we can expand the reflection coefficient as
\begin{equation}
\label{eq:pt_ref2}
    r=r'= -\frac{i\pi \epsilon}{\sinh(\pi \omega r_s)} + \mathcal{O}(\epsilon^2).
\end{equation}
Expanding the leading order in $\epsilon$ also for small values of $\omega r_s$ one finds
\begin{equation}
    r=r'= -\frac{i \epsilon}{\omega r_s} +\mathcal{O}((\omega^2 r_s^2)),
\end{equation}
where this result holds only while keeping $|r|,|r'|\ll 1$.\\

For the Schwarzschild case the potential for axial gravitational perturbations reads
\begin{equation}
    V_0= \left(1-\frac{r_s}{r}\right)\left[\frac{\ell(\ell+1)}{r^2}-\frac{3r_s}{r^3}\right],
\end{equation}
and from  Ref.~\cite{transfer_will_wkb_PhysRevD.35.3621} we get that, in a WKB approximation,

\begin{equation}
    \frac{1}{t'}=\left( \frac{2\pi i e^{i\pi \nu}}{\Gamma(-\nu)\Gamma(\nu+1)} + e^{2\pi i \nu} \right)^{-1}\frac{R^{-2}(2\pi)^{1/2}}{\Gamma(-\nu)}
\end{equation}
with
\begin{equation}
    r = r' =\frac{e^{i\pi \nu} R^2 \,\Gamma(-\nu)}{(2\pi)^{1/2}},
\end{equation}
where 
\begin{equation}
    R= \left( \nu +\frac 12\right)^{\left(\nu + \frac 12\right)/2} e^{-\left(\nu + \frac 12\right)/2}.
\end{equation}
We used the definition from Ref.~\cite{Schutz:1985km}
\begin{equation}
    \nu +\frac 12 = -i \frac{\omega^2 - \tilde{V}_{0,\ell}}{\sqrt{-2\tilde{V}_{0,\ell}''}}
\end{equation}
where $\tilde{V}_{0,\ell}$ and $\tilde{V}_{0,\ell}''$ are respectively the potential $V_0$ evaluated at the maximum and its second derivative evaluated at the maximum, for a specific $\ell$. Furthermore $\nu\in \mathbb{N}$ if we pick the poles of $t'$ to get the QNM solutions.
Using the WKB approximation, each pole $\nu = n$ gives the frequency
\begin{equation}
\label{eq:wkb_frequencies}
    \omega_{\ell,0}^2 = \tilde{V}_{0,\ell} - i\left(n+\frac{1}{2}\right)\sqrt{-2\tilde{V}_{0,\ell}''}.
\end{equation}
The residue of the reflection coefficient around $\nu=n$ is, for $\ell=2$, 
\begin{equation}{\rm Res} \, r |_{\ell=2}=-\frac{\left(-1\right)^{2n } e^{-\frac12-n-\frac1{12+24n}}\left(\frac12+n\right)^{n}\sqrt{1+2n}}{\sqrt{\pi}n!}.\end{equation}
Picking the least damped solutions $n=0,1$ we get
\begin{equation}
    {\rm Res}\; r|_{n=0,\ell=2} = -\frac{1}{e^{7/12}\sqrt{\pi}}\simeq -0.314,\,\,\,{\rm Res}\; r|_{n=1,\ell=2} = -\frac{3 \sqrt{\frac{3}{\pi }}}{2 e^{55/36}}\simeq -0.320.
\end{equation}

\section{Two-delta example}
\noindent
In this section we provide a toy model to understand the various regimes in which we evaluated the effect given by the presence of the bump. In this simplified model we consider the two-delta potential
\begin{equation}
    V = 2 \beta \delta(r_*) + 2\epsilon \beta \delta(r_* - c), 
\end{equation}
where $\beta$ is real and can be thought of order $\beta \sim r_s^{-1}$. The main advantage of this model is that the transmission and reflection coefficients are known,
\begin{equation}
    t_0(\omega) = \frac{1}{1+ i \beta/\omega}, \quad r_0(\omega) = \frac{-i\beta/\omega }{1+ i \beta/\omega},
\end{equation}
and the same for $t_1$ and  $r_1$ just by replacing  $\beta$ by $\epsilon \beta$, 

\begin{equation}
    t_1(\omega) = \frac{1}{1+ i\epsilon \beta/\omega}, \quad r_1(\omega) = \frac{-i\epsilon\beta/\omega }{1+ i \epsilon\beta/\omega}.
\end{equation}
From the transmission coefficients we can immediately see that the two typical wavelengths related to their poles are respectively of order
\begin{equation}
    \omega_0 = -i \beta, \quad \omega_b = -i \epsilon\beta.
\end{equation}
To find the frequency of the two-delta system we need to solve Eq.~(\ref{eq:t_tot}), remembering that $r' = - r^* t/t^*  $ because both the potentials are symmetric under time reversal. Therefore  Eq.~(\ref{eq:t_tot}) becomes
\begin{equation}
\label{eq:deltas_transmission}
    e^{-2i\omega c} = -\frac{r_0^* t_0}{t_0^*}r_1= -\frac{\epsilon\beta^2/|\omega|^2 }{(1+ i \beta/\omega)(1+ i \epsilon\beta/\omega)},
\end{equation}
which can be solved numerically. However we perform two limits to understand the behaviour of the solution found and gain insight on the original problem.

The first limit is for a spectrum close to the unperturbed one. In the limit $\epsilon\to 0$ we have that the solution at leading order in $\epsilon$ can be written as
\begin{equation}
    \omega = \omega_0 + \epsilon \omega_1,
\end{equation}
and substituting in the equation above while keeping only the first order in $\epsilon$ we get
\begin{equation}
    e^{-2i\omega_0 c}\simeq -\frac{\beta^2/|\omega_0|^2 }{\omega_1/\omega_0 (1+ i \epsilon\beta/\omega_0)} \quad {\rm or}\quad \omega_1 \simeq i \beta e^{2\beta c}.
\end{equation}
The corresponding  frequency is
\begin{equation}
    \omega = \omega_0 + \epsilon \omega_1 = -i\beta \left( 1-\epsilon\,  e^{2\beta c}\right), 
\end{equation}
which grows exponentially with $c$ as expected.
This result is valid for all the values of $c$, until the point the perturbation reaches the same size of the original frequency, i.e. until $c=c_*$ such that
\begin{equation}
    \epsilon \, e^{2\beta c_*}\sim 1.
\end{equation}
However, even when the value  $c_*$ is reached,  we still have 
\begin{equation}
    \epsilon \beta c_* \sim \epsilon \log(1/\epsilon) \ll 1,
\end{equation}
for typical values $\epsilon\sim (10^{-5} \div 10^{-3})$ and $\beta c\sim (10\div10^2)$ used in this problem. Therefore we can identify a second region for $c$ where $c\gsim c_*$ where
\begin{equation}
    \epsilon \, e^{2\beta c_*} \gsim 1, \quad \epsilon \beta c_*\ll 1.
\end{equation}
In this region our intuition is that the wave decays and looses most of its amplitude before reaching the second peak, therefore getting reflected and slowly forming a quasi-bound state between the peaks. This insight leads us to expect frequencies of the order $\omega \sim 1/c$.

From Eq.~(\ref{eq:deltas_transmission}) it is possible to notice that neither pole is picked given our estimate for the frequency. Consistently with our estimate, in the denominator we can approximate
\begin{equation}
    \frac{1}{1+ i \beta/\omega}\sim   \frac{1}{i \beta/\omega},\quad {\rm and} \quad \frac{1}{1+ i \epsilon \beta/\omega}  \sim 1
\end{equation}
and we need to solve the simpler equation
\begin{equation}
    e^{-2i\omega c} = i\frac{\epsilon \beta \omega}{|\omega|^2}.
\end{equation}
Expanding the equation and writing $\omega = \omega_R + i \omega_I$, the only solution is  $\omega_R=0$, giving
\begin{equation}
    e^{2  \omega_I c}=-\frac{\epsilon \beta }{\omega_I}
\end{equation}
which has two solutions, one with $\omega_I \sim 0$ and the other which can be approximated as
\begin{equation}
    \omega_I \sim -\frac{\log(\epsilon \beta c)}{2c},
\end{equation}
giving a frequency written as
\begin{equation}
    \omega \sim -\frac{i}{2c}\log(\epsilon \beta c).
\end{equation}
There are therefore two regimes, depending on the position of the bump. At first, the frequency is close to the original pole and it migrates, still being close to the original value. This changes when the perturbation become $\mathcal{O}(1)$, giving rise to an inverse distance regime, similar to a  confined metastable particle in a potential well.  

\section{The effective metric}
\noindent
For completeness, in this section we reconstruct the metric associated with a bumpy potential. For  simplicity we will consider the case of a scalar perturbation, although the argument can be generalized.
Consider the equation of motion of a massless scalar field in the Schwarzschild spacetime to which a small potential bump is added away from   the location of the photon ring

\begin{equation}
\label{eqsscalar}
\left[\partial_{r_*}^2+\omega^2-V^\epsilon_\ell(r)\right] \phi_{\ell}(r)=0,  \end{equation}
where the 
 potential reads
\begin{eqnarray}
V^\epsilon_\ell(r)&=&V_0(r)+\epsilon V_1(r),\nonumber\\
V_0(r)&=&\left(1-\frac{r_s}{r}\right)\left[\frac{\ell(\ell+1)}{r^2}+\frac{r_s}{r^3}\right],\nonumber\\
V_1&=&
 \frac{2}{r_s}\delta(r-c).  
\end{eqnarray}
The Dirac delta in the potential is located at a distance $c\gg $ than the location $r_g$ of the peak of the main potential (where the photon ring resides), has an amplitude ${\cal O}(\epsilon)$ and mimics a more realistic bump with a width ${\cal O}(r_s)\ll c$.

Now, we wish to answer the following question: what is the effective metric  which gives Eq.~(\ref{eqsscalar})? For large distances of the bump, $r_*\simeq r$, and writing  the effective metric as

\begin{equation}
\label{metric}
 {\rm d} s^2=-A(r) {\rm d}t^2+\frac{1}{A(r)}{\rm d}r^2+r^2{\rm d}\Omega^2,  
\end{equation}
the equation of motion of the scalar field becomes

\begin{equation}   \left[\partial_{r_*}^2+\omega^2-A\frac{\partial_r A}{r}-A\frac{\ell(\ell+1)}{r^2}\right]\phi_\ell=0.
\end{equation}
Setting

\be
 A(r)=A_0(r)+\epsilon A_1(r),   
\ee
we find

\begin{eqnarray}
A_0(r)&=&1-\frac{r_s}{r},\nonumber\\
\frac{A_0}{r}\partial_rA_1&=&V_1-\frac{A_1}{A_0}V_0.
\end{eqnarray}
We can integrate the last equation  to find

\begin{eqnarray}
A_1(r)&=&\frac{K r}{(r-r_s)^{\ell(\ell+1)+1}} \nonumber\\
&+&2\frac{ c\, r(c-r_s)^{\ell(\ell+1)}}{r_s(r-r_s)^{\ell(\ell+1)+1}}\theta(r-c),
\end{eqnarray}
where $K$ is an integration constant, which we set to zero so that the boundary condition is $A_1(r_s\ll r\ll  c)=0$.

For $r$ and $c\gg r_s$, the additional piece in the metric reduces to

\be
 A_1(r)\simeq \frac{2 c}{r_s}\left(\frac{c}{r}\right)^{\ell(\ell+1)}\theta(r-c).
\ee
The effective metric  looks as follows: at distances $r_g\ll r\ll  c$ it is the usual Schwarzschild metric; at $r=c$ it has a bump ${\cal O}(2\epsilon c/r_s)$ and then it decays as $ r^{-\ell(\ell+1)}$.

Within the spacetime with metric (\ref{metric}), null geodesics are determined  by the trajectories which move along the equatorial plane such that

\be
-A\dot t^2+\frac{1}{A}\dot r^2+r^2\dot\phi^2=0,
\ee
where the dots indicate differentation with respect to the affine parameter and $\phi$ is the azimuthal angle. Because of the spherical symmetry, one has $\dot\phi^2=L^2/r^4$, where $L$ is the angular momentum. Similarly, stationarity gives $\dot t^2=E^2/A^2$, where $E$ is the conserved energy. The equation of motion can be written as

\be
\label{eq:potential}
\dot r^2=-V(r)=-A\left(-\frac{E^2}{A}+\frac{L^2}{r^2}\right).
\ee
A circular orbit at a given radius $r_c$ exists if 

\be
V(r_c)=V'(r_c)=0,
\ee
where the prime indicates differentation with respect to radial coordinate. These conditions impose, respectively

\begin{eqnarray}
    \frac{E^2}{L^2}&=& \frac{A_c}{r^2_c},\\
    \label{con}
    1&=&r_c\frac{A'_c}{2A_c},
\end{eqnarray}
where the subscript ${}_c$ means that the quantity in question is evaluated at the radius $r_c$ of the  circular null geodesic. 

In the problem at hand, there are three locations  of $r_c$: the maxima of $A_0$ and $A_1$ (unstable orbits) and the minimum located close to $c$ where $A_0\simeq \epsilon A_1$ (stable orbit). Since the angular velocity at these values reads 

\be
\omega_c=\frac{A_c}{r_c},
\ee
for the stable orbit we find $\omega_c(r_c\simeq c)\simeq 1/c$, reproducing the scaling for the quasi-bound state.

\end{document}